\documentclass[preprint,%
tightenlines,%
superscriptaddress,%
floats,prd,eqsecnum,nobibnotes%
,nofootinbib,aps,12pt]{revtex4}
\usepackage{amsmath, amssymb}

\newif\ifeprint
\eprinttrue

\newcommand\abs[1]{\ensuremath{\left|{#1}\right|}}
\newcommand\com[2]{\ensuremath{\left[{#1},{#2}\right]}}
\DeclareMathOperator\Tr{Tr}
\newcommand\cfd{conformal\ }
\newcommand\hepth[1]{\eprint{{\ifeprint\tt\fi hep-th/#1}}}
\newcommand\hepph[1]{\eprint{{\ifeprint\tt\fi hep-ph/#1}}}

\newcommand\oldappendix{This will give an error if \backspace oldappendix
   already exists.}
\let\oldappendix\appendix
\renewcommand\appendix{\oldappendix%
   \renewcommand\theequation{\thesection.\arabic{equation}}}

\newcommand{\beq}{\begin{equation}}
\newcommand{\eeq}{\end{equation}}
\newcommand{\beqa}{\begin{eqnarray}}
\newcommand{\eeqa}{\end{eqnarray}}
\newcommand{\be}{\begin{equation}}
\newcommand{\ee}{\end{equation}}
\newcommand{\bea}{\begin{eqnarray}}
\newcommand{\eea}{\end{eqnarray}}
\newcommand{\bA}{\begin{array}}
\newcommand{\eA}{\end{array}}
\newcommand{\bc}{\begin{center}}
\newcommand{\ec}{\end{center}}
\newcommand{\al}{\alpha}

\newcommand{\ra}{\rightarrow}
\newcommand{\del}{\partial}
\newcommand{\p}{\partial}
\newcommand{\ie}{{\em i.e.}}
\newcommand{\eg}{{\em e.g.}}
\newcommand{\Z}{\mathbb{Z}}

\newcommand{\C}{\mathbb{C}}

\newcommand{\Nf}{${\mathcal N}{=}4$}

\newcommand{\vx}{{\vec x}}

\newcommand{\vk}{{\vec{k}}}

\newcommand\tF{{\tilde{F}}}
\newcommand\tA{{\tilde{A}}}

\newcommand{\ben}{\begin{eqnarray}\displaystyle}
\newcommand{\een}{\end{eqnarray}}

\begin{document}

\ifeprint
\setlength{\baselineskip}{1.2\baselineskip} 
\fi


\title{\vspace*{\fill}\LARGE Cosmologies with Null Singularities \\
and their Gauge Theory Duals}

\author{\large Sumit R. Das}\email{das@pa,uky,edu}
\affiliation{Department of Physics and Astronomy, 
      University of Kentucky, 
      Lexington, Kentucky \ 40506 \ U.S.A. \vspace*{.5\baselineskip}}
\author{\large Jeremy Michelson}\email{jeremy@pa,uky,edu}
\affiliation{Department of Physics, 
      The Ohio State University,
      1040 Physics Research Building, 
      191 West Woodruff Avenue, 
      Columbus, Ohio \ 43210-1117 \ U.S.A. \vspace*{.5\baselineskip}}
\author{\large K. Narayan}\email{narayan@theory,tifr,res,in}
\author{\large Sandip P. Trivedi} \email{sandip@theory,tifr,res,in}
\affiliation{Department of Theoretical Physics, 
Tata Institute of Fundamental Research, 
Homi Bhabha Road, Colaba, 
Mumbai - 400005, INDIA.
\vspace*{\fill}} 

\begin{abstract}
\vspace*{\baselineskip}

We investigate backgrounds of Type IIB string theory with null
singularities and their duals proposed in \hepth{0602107}. The
dual theory is a deformed ${\mathcal N}=4$ Yang-Mills theory in 3+1
dimensions with couplings dependent on a light-like direction.  We
concentrate on backgrounds which become $AdS_5 \times S^5$ at early
and late times and where the string coupling is bounded, vanishing at
the singularity. Our main conclusion is that in these cases the dual
gauge theory is nonsingular.  We show this by arguing that there
exists a complete set of gauge invariant observables in the dual gauge
theory whose correlation functions are nonsingular at all times.  The
two-point correlator for some operators calculated in the gauge theory
does not agree with the result from the bulk supergravity solution.
However, the bulk calculation is invalid near the singularity where
corrections to the supergravity approximation become important.  We
also obtain pp-waves which are suitable Penrose limits of this general
class of solutions, and construct the Matrix Membrane theory which
describes these pp-wave backgrounds.

\end{abstract}
\preprint{\parbox[t]{10em}{\begin{flushright}
TIFR/TH/06-29 \\
UK/06-12 \\ {\tt hep-th/0610053} 
\end{flushright}}}

\maketitle

\ifeprint
\tableofcontents
\fi


\section{Introduction and Summary}

Time dependent backgrounds, particularly those which contain
space-like or null singularities, are amongst the most
poorly understood aspects of string theory. This problem has been attacked
from various viewpoints for a long time. These include
perturbative string theory \cite{horowitzsteif, bhkn, cornalbacosta, 
lms1, lms2, ckr, keshav0302, grs, bp0307, bdpr}, 
open and closed string tachyon condensation \cite{tachyons},
particularly those which lead to space-like or null singularities
\cite{Aharony:2002cx, McGreevy:2005ci, Silverstein:2005qf,
Horowitz:2006mr}.

Recently there have been several attempts to attack this problem using
holographic duals of various kinds.  These include matrix model
formulations of noncritical string theory
\cite{karczstrom,Karczmarek:2004ph,Das:2004hw, daskarcz0412, dp,
das0503} (which also involve closed string tachyon condensation),
Matrix Theory duals of backgrounds with null linear dilatons
\cite{csv0506, li0506, berkooz0507, lisong0507, Hikida:2005ec,
dasmichel0508, Chen:2005mg, she0509, Chen:2005bk, Ishino:2005,
robbinssethi0509, KalyanaRama:2005uw, hikidatai0510, Li:2005ai,
Craps:2006xq, dmichelson2, sethi0603, chen0603, ohta0603, nayak0604,
ohta0605, craps0605, nayak0605}, and use of the AdS/CFT correspondence
\cite{keshav0302, Hertog:2004rz, Hertog:2005hu,chuho0602,dmnt0602,linwen}. 
It has been suspected for a long time that the usual notions of
space-time break down near singularities.
These holographic approaches attempt to make
this idea concrete by replacing usual dynamical space-time by a more
fundamental structure provided typically by a gauge theory in lower
number of dimensions.

In a previous paper, \cite{dmnt0602}, we reported on the construction
of a family of solutions in Type IIB string theory. The solutions are
either time-dependent or depend on a light-like
coordinate\footnote{Below, we will loosely refer to backgrounds which
depend on a light-like coordinate as null backgrounds.}, and often
have singularities which are space-like or null respectively.  The
solutions can be thought of as deformations of the well known $AdS_5
\times S^5$ solution.  The dilaton and axion are also excited in these
solutions, and in some of them the dilaton remains weakly coupled
everywhere, including the singularity.  Similar solutions were studied
in \cite{chuho0602,linwen}.

In this note we continue our study of these backgrounds. In
particular, we focus on the null backgrounds, with a {\em weakly} coupled
dilaton.  The metric and dilaton in these solutions take the form,
\be
\label{igeom}
ds^2=(\frac{r^2}{R^2}){\tilde g}_{\mu\nu}(X^+) dx^\mu dx^\nu +
(\frac{R^2}{r^2})dr^2 + R^2 d\Omega_5^2, \qquad \Phi=\Phi(X^+),
\ee
 with the four-dimensional metric,
\be
\label{imetpar}
d{\tilde s}^2   =   {\tilde g}_{\mu\nu}dx^\mu dx^\nu
=e^{f(X^+)}(-2dX^+dX^-+dx_2^2+dx_3^2). 
\ee
Singularities can arise when the conformal factor $e^f$ vanishes. 
When this happens, in the solutions of interest, $e^\Phi$
vanishes at the singularity. Also, asymptotically as 
$X^+\rightarrow -\infty$, $e^f$ and the dilaton $\Phi$ go to a 
constant, so that these solutions asymptote  to the familiar 
$AdS_5 \times S^5$ solution. This is in contrast with the kind
of backgrounds studied using some other approaches, where the bulk
string coupling is typically {\em large} near the singularity.

In this paper, we argue that these backgrounds have a dual description
as the ${\mathcal N}=4$ Super Yang-Mills theory living in a $3+1$
dimensional spacetime with metric ${\tilde g}_{\mu\nu}$ and with a
varying Yang-Mills coupling constant $g_\text{YM}^2=e^\Phi$. The Yang-Mills
theory starts in the ${\mathcal N}=4$ vacuum state as $X^+ \rightarrow
-\infty$ and we want to understand the time evolution of this system
as we approach the singularity at $X^+ = 0$.

Our main conclusion is that the gauge theory is nonsingular.  By this
we mean that we can find a complete set of gauge invariant operators
whose correlation functions are nonsingular even when the bulk
geometry has a singularity. This happens because of two features of
our null solutions. First, the metric, eq.~(\ref{imetpar}), is conformally
flat, and one finds that due to the light-like dependence of the
conformal factor $e^f$, the conformal anomaly of the gauge theory in
this background vanishes.  Secondly, the dilaton and hence the
Yang-Mills coupling becomes arbitrarily weak near $X^+ = 0$.

In more detail, we carry out the analysis in two parts. 
First, we neglect the varying dilaton and study the gauge theory in the 
presence of the nontrivial conformal factor $e^f$. 
Using the fact that the conformal anomaly vanishes, we then argue that
if the operators, ${\mathcal O}_i(x),$ have 
conformal dimensions $\Delta_i$, 
their correlation functions satisfy  the relation,  
\be
\label{igenindt}
\langle e^{\frac{f(x_1)\Delta_1}{2}} {\mathcal O}(x_1)
e^{\frac{f(x_2) \Delta_2}{2}}
{\mathcal O}(x_2) \cdots
e^{\frac{f(x_n) \Delta_n}{2}} {\mathcal O}(x_n)\rangle_{[e^f \eta_{\mu\nu}]}=
\langle{\mathcal O}(x_1) {\mathcal O}(x_2) \cdots {\mathcal O}(x_n)
\rangle_{[\eta_{\mu\nu}]}.
\ee
Here the left hand side is calculated in the theory with the metric, 
 ${\tilde g}_{\mu\nu}= e^f \eta_{\mu\nu}$, while the right hand side is 
calculated with the flat metric. Both correlation functions are calculated 
in the vacuum of the ${\mathcal N}=4$ theory, which is conformally 
invariant\footnote{Note that we are using the phrase ``conformal 
invariance" in the sense of Weyl invariance, \ie\ a position dependent
rescaling of the metric without any coordinate transformation. Therefore, 
the conformal dimensions $\Delta$ which appear in (\ref{igenindt}) can
be distinct from usual dimensions of operators under \eg\ rescaling of
coordinates.}.

In the second part of the analysis, we include the effects of the
varying coupling constant due to the time dependent bulk dilaton. 
The important point here is that for the supergravity solutions of
interest, $e^\Phi$ is small everywhere and {\em vanishes} at
the singularity. Thus, its effects can be controlled. We show that the
varying dilaton, due to its null dependence, does not give rise to any
particle production in the interaction picture. Near the singularity,
since $e^\Phi$ becomes small and vanishes, the gauge field,
$A_\mu$, with quadratic terms,
\be
\label{iact}
S_\text{GF}
=-\frac{1}{4}\int d^4x~e^{-\Phi} \Tr[F_{\mu\nu}F^{\mu\nu}],
\ee
has singular kinetic terms and is not a well defined variable. 
Working in light cone gauge, $A_-=0$, a well defined variable which has 
canonical quadratic terms is given by 
\be
\label{iatilde}
{\tilde A}_\mu=e^{-\frac{\Phi}{2}}A_\mu.
\ee
For these null backgrounds, we argue that gauge invariant operators 
made from the ${\tilde A}$ variables are nonsingular at the 
singularity\footnote{The conformal dimension of $A_\mu$ is zero. For 
example, the action of the gauge theory is  classically Weyl invariant, 
provided $A_\mu$ has vanishing conformal weight; see 
\eg~\cite[p.~448]{Walda}. This means the conformal weight of 
${\tilde A}_\mu$ also vanishes and so the ${\tilde A}_\mu$ variables
do not require any dressing.}.  This is because the gauge coupling
vanishes at $X^+=0$ so that coupling effects do not destroy the
nonsingular nature of the propagator and also render higher point
functions nonsingular.  Generically, these operators have supergravity
duals which are not local excitations in the bulk.
 
At the singularity, $e^{\Phi}$ vanishes and the 't Hooft
coupling $g_\text{YM}^2N$ in the gauge theory goes to zero. This suggests 
that $\alpha'$ corrections become important near the singularity, since 
for the $AdS_5\times S^5$ duality we have $\al'\sim \frac{1}{
g_\text{YM}^2N}$ . To understand this issue, we consider the bosonic
part of the worldsheet action for a fundamental string in this
class of background and fix the light cone gauge. The gauge fixed
action clearly shows that near the singularity all the oscillator modes
of the string become very light. We have not performed a detailed
analysis of the worldsheet theory to examine whether perturbative
string theory can be consistently defined  in this background.

Our results strongly suggest that using the nonsingular description
of the gauge theory we can extend the bulk spacetime past the
singularity. In some sense, the singularity therefore appears to be a
problem caused by a wrong choice of dynamical variables. In
particular, variables which are natural and local from the bulk ten
dimensional supergravity point of view are {\em not} well behaved at the
singularity.
However, a correct choice of variables, which is transparent in the
dual gauge theory, ``resolves'' this singularity.  We find that the
correct description is obtained by appropriately continuing the metric
and the dilaton.  The resulting spacetime is asymptotically
$AdS_5\times S^5$ in the far future, as $X^+ \rightarrow +\infty$, with
a constant dilaton.

Finally we perform the Penrose limit of our class of solutions by
zooming in on  a suitable null geodesic and obtain the resulting
pp-wave solutions. 
The maximally supersymmetric
type IIB pp-wave with one compact null direction and an additional
compact space-like direction 
is a unique background for which two kinds of
holographic duality are understood: the holographic duality to a
certain sector of Yang-Mills theory in 3+1 dimensions
\cite{Mukhi:2002ck, Bertolini:2002nr, DeRisi:2004bc} along the lines
of \cite{Berenstein:2002jq}, and the duality to DLCQ Matrix theory
\cite{Gopakumar:2002dq, Michelson:2004fh} along the lines of
\cite{Banks:1996vh, Banks:1996my, Motl:1997th}. In \cite{dmichelson2}
type IIB pp-waves with a dilaton which is linear in a null time was
considered as a model of a null big bang. It would be of interest to
relate the insight gained from a AdS/CFT perspective to that obtained
from a DLCQ matrix perspective and possibly establish a precise
relationship between the two. Using the methods outlined in
\cite{dmichelson2} we write down the 2+1 dimensional Yang-Mills theory
which is dual to string theory in the pp-waves which 
arise from the solutions considered in this paper. A more detailed analysis 
of this Matrix Membrane theory near the singularity is left for the future. 

Some of our analysis and conclusions contain substantial overlap with
\cite{chuho0602, linwen}. The supergravity solutions were found and their
conjectured gauge theory duals identified in \cite{chuho0602,
dmnt0602, linwen}.  Furthermore \cite{chuho0602} calculated the bulk two-point
function for the null backgrounds (which agrees with our calculation
in sec.~V). Some discussion of dressed correlators and the suggestion
that the gauge theory is in fact nonsingular is also contained in
\cite{chuho0602}.

This paper is structured as follows. The supergravity solutions are
reviewed in section~\ref{sec:soln}.  The gauge theory duals are
identified in section~\ref{sec:dual}, and extensively analysed in
section~\ref{sec:analysis}.  Some two-point functions are calculated
in the bulk in section~\ref{sec:green}.  Section~\ref{worldsheet}
concerns the bosonic part of the worldsheet action for a string moving
in this class of backgrounds. In Section~\ref{sec:matrix} we perform a
Penrose limit and write down the Matrix membrane theory action.

\section{Review of Supergravity Solutions} \label{sec:soln}

The  solutions we are interested in are discussed in section 2 of 
\cite{dmnt0602}.%
\footnote{See 
also \cite{keshav0302, Ishino:2005, keshav0501, ohta0605} for 
time-dependent supergravity solutions in an M-theory context.}
The ten-dimensional Einstein frame metric and the 
dilaton are,
\begin{subequations} \label{geomdil}
\begin{align}
\label{geom}
ds^2&=(\frac{r^2}{R^2}){\tilde g}_{\mu\nu} dx^\mu dx^\nu +
(\frac{R^2}{r^2})dr^2 + R^2 d\Omega_5^2, \\
F_{(5)}&=R^4(\omega_5 + *_{10} \omega_5), \\
\label{formdil}
\Phi&=\Phi(x^\mu).
\end{align}
\end{subequations}
This is a solution of the equations of motion as long as the four-dimensional 
metric, ${\tilde g}_{\mu\nu}$, and the dilaton, $\Phi$, are only dependent 
on the four coordinates, $x^\mu, \mu =0,1,2,3$, and satisfy the conditions, 
\begin{subequations}\label{conds}
\begin{gather}
{\tilde R}_{\mu\nu} = \frac{1}{2} \partial_\mu\Phi \partial_\nu\Phi,
\\
 \partial_\mu (\sqrt{-\det({\tilde g})}\ {\tilde g}^{\mu \nu}
\partial_\nu \Phi) = 0,
\end{gather}
\end{subequations}%
where ${\tilde R}_{\mu\nu}$ is the Ricci curvature of the metric 
${\tilde g}_{\mu\nu}$. 
In eq.~(\ref{geomdil}), $d\Omega_5^2$ is the volume element and 
$\omega_5$ is the volume form of the unit five sphere. 

Of particular interest in this paper is the case where 
${\tilde g}_{\mu\nu}$ is conformally flat and where  both 
 ${\tilde g}_{\mu\nu}$ and $\Phi$ only depend on one light-like
coordinate which we take to be $X^+$. 
In this case, the four-dimensional spacetime background takes the form, 
\begin{subequations} \label{metdil}
\begin{align}
\label{metpar}
d{\tilde s}^2 = {\tilde g}_{\mu\nu}dx^\mu dx^\nu
&=e^{f(X^+)}(-2dX^+dX^-+dx_2^2+dx_3^2), \\
\label{vardilp}
\Phi & = \Phi(X^+). 
\end{align}
\end{subequations}
The conditions, eq.~(\ref{conds}), then require, 
\be
\label{condnull}
\frac{1}{2}(f')^2 - f'' = \frac{1}{2}(\del_+ \Phi)^2.
\ee
Generically a non-flat metric ${\tilde{g}}_{\mu\nu}$ as in
(\ref{geom}) introduces curvature singularities at the Poincare horizon
$r=0$. It can be easily checked, however, that these singularities are
absent for such null backgrounds. The only possible singularities appear
at values of $X^+$ where time-like or null geodesics entirely lying 
in the $X^\mu$ subspace end or begin at finite affine parameters.

An important case, prototypical of the kind of example we have in mind
throughout this paper, is obtained by taking
\be
\label{excf}
e^f  =  \tanh^2X^+.
\ee 
Then
\be
\label{nullsolntanh}
d{\tilde s}^2 = \tanh^2 X^+ (-2dX^+ dX^- + dx_2^2 + dx_3^2),
\qquad e^{\Phi}=g_s \left|\tanh \frac{X^+}{2}\right|^{\sqrt{8}}.
\ee
In this example\footnote{Note that a solution with $e^{f(X^+)}=\tanh^2(QX^+)$ 
is equivalent to eq.~(\ref{nullsolntanh}), by setting to unity the 
dimensionful scale $Q$ using the symmetry
$X^+\ra\lambda X^+ ,\ Q\ra \frac{Q}{\lambda},\ X^-\ra\frac{X^-}{\lambda}$ , 
which is special to these null solutions.\label{ft:rescale}},
in the far past and future
as $X^+\rightarrow\pm\infty$, the four-dimensional spacetime becomes 
asymptotically flat and the dilaton goes to a constant. By choosing 
$g_s$ to be small, the string coupling $e^\Phi$
can be made small everywhere in the 
spacetime\footnote{Up to a shift or rescaling (footnote~\ref{ft:rescale})
of $X^+$, the only nontrivial ($e^f \neq \text{const}$) solution with an 
everywhere constant dilaton is $e^f=\frac{1}{(X^+)^2}$, which can be 
seen to be flat space by the coordinate transformation
\begin{align*}
x_2&=X^+ Y_2 , & x_3&=X^+ Y_3 , &
X^-&=Y^--X^+(Y_2^2+Y_3^2) , & X^+&=-\tfrac{1}{Y^+}.
\end{align*}}.
In the example, eq.~(\ref{nullsolntanh}), $e^\Phi$ is not 
analytic at $X^+=0$. It satisfies the equation, eq.~(\ref{condnull}), for 
$X^+>0$ and $X^+<0$, and is continuous at $X^+=0$.  

For the metric, eq.~(\ref{geom}) obeying
eq.~(\ref{metpar}), the affine parameter
$\lambda$, for a null geodesic moving along a trajectory with 
$X^-, x^2,x^3$ constant, is given by
\be
\label{deflambda}
\lambda = \int e^{f(X^+)} dX^+.
\ee
The tangent vector along this geodesic is 
\be
\label{defxi}
\xi=\xi^{\mu}\del_{\mu}\equiv \frac{\del}{\del\lambda}=
\biggl(\frac{d\lambda}{dX^+}\biggr)^{-1} \frac{\del}{\del X^+}
= e^{-f} \frac{\del}{\del X^+}, \qquad \xi^{\mu}\xi_{\mu}=0.
\ee

For the ten-dimensional metric, eq.~(\ref{geom}), the Ricci scalar,
$R^A{_A}$ and the invariants, $R_{AB}R^{AB}, R_{ABCD}R^{ABCD},$ are all
constant and independent\footnote{ Here the indices $A,B \cdots,$
take values in the tangent space of $x^\mu,r$, eq.~(\ref{geom}). The
statement is also true if they range over the full ten-dimensional
tangent space including the $S^5$.} of $r, X^+$.  In contrast, the
curvature invariant, for the metric, eq.~(\ref{geom}),
$R_{ab}\xi^a\xi^b$, ($\xi^a$ is the component of the tangent vector,
eq.~(\ref{defxi})) is,
\be
\label{affcurv}
R_{\lambda \lambda} =R_{++} \left(\frac{dX^+}{d\lambda}\right)^2 
= \left(\frac{1}{2}(f')^2-f^{''}\right) e^{-2f}.
\ee
For a suitably chosen $f$ this can blow up when $e^f \rightarrow 0$,
and the resulting value of $\lambda$, eq.~(\ref{deflambda}), can be finite.
This results in a singularity which occurs at finite affine time. 

For example, in the case, eq.~(\ref{nullsolntanh}),
$R_{++}=\frac{4}{\sinh^2 X^+}$ , so that
$R_{\lambda\lambda}=\frac{4}{\sinh^2X^+ \tanh^4X^+}$ ,
 showing a curvature singularity at
$X^+\ra 0$ , with $e^{\Phi}$ becoming arbitrarily small there.%
\footnote{The string frame
curvature, $R_{\lambda \lambda}$ blows up at the singularity in this case
as well.} One can see in this example that the singularity occurs at finite
affine time.

It is worth understanding the resulting singularity better. 
Consider  null geodesics with $X^+$ varying and all other coordinates fixed. 
Take two such nearby geodesics displaced along the $x^i, i=2,3,$ directions. 
Following \cite[p.~47]{Walda}, the relative acceleration of these geodesics is
\be
\label{relacc}
a^i=-R{_{+i+}}^i(\xi^+)^2= -\frac{1}{2}\left(\frac{1}{2}(f')^2-f''\right) 
e^{-2f},
\ee
where $\xi$ is defined in eq.~(\ref{defxi}).
Up to a sign, this is exactly $R_{\lambda \lambda}$, eq.~(\ref{affcurv}). 
We see that there is a diverging compressional tidal force as the 
singularity is approached. Another way to see this is to  calculate the 
physical distance between two nearby null geodesics of this type. For
two geodesics displaced along $x^i$, the physical distance is  
\be
\label{phydist}
\Delta=\frac{e^{f/2}}{z} \sqrt{(x_i)^2},
\ee
satisfying the equation 
\be
\label{eqpd}
\frac{d^2\Delta}{d\lambda^2}=-\frac{1}{2}\left(\frac{1}{2} (f')^2 -f''\right) 
\Delta.
\ee 
The first term on the right-hand side (rhs)
agrees with $a^i$, eq.~(\ref{relacc}), and 
$R_{\lambda\lambda}$, eq.~(\ref{affcurv}), showing once again that the 
compressional tidal force diverges at the singularity.  

It is also worth mentioning that the example, eq.~(\ref{nullsolntanh}), can 
be regarded as a limiting case of a one-parameter family of solutions, with 
the conformal factor, 
\be
\label{onep}
e^f=(|\tanh X^+| + \epsilon)^2,
\ee
and the dilaton given by solving eq.~(\ref{condnull}). For small $\epsilon$, 
the dilaton deviates from its value, eq.~(\ref{nullsolntanh}), only close to 
$X^+=0$. At $X^+=0$, $e^\Phi\sim g_s (\epsilon)^{\sqrt{8}}$, so that the 
string coupling is non-vanishing but small. Note that in eq.~(\ref{onep})
the metric
is continuous but non-analytic at $X^+=0$ and the first derivative of 
$e^f$ has a finite discontinuity
there.
However, the curvature component, $R_{++}$, is continuous and nonsingular at $X^+=0$,
and the affine parameter, $\lambda$, eq.~(\ref{deflambda}), is also continuous there. 

We close by noting that the null backgrounds, eqs.~\eqref{geomdil}
and~(\ref{metdil}), preserve 
8 supersymmetries.

\section{The Dual Field Theory} \label{sec:dual}

In this section we argue that the backgrounds discussed above are dual
to the ${\mathcal N}=4$ Super Yang-Mills theory in a $3+1$ dimensional
spacetime with a varying background metric and a varying gauge
coupling.  The background metric is ${\tilde g}_{\mu\nu}$,
eq.~(\ref{geom}), and the gauge coupling is given in terms of the
dilaton, eq.~(\ref{formdil}), by, $g_\text{YM}^2=e^\Phi$.

We now review  the evidence in support of this claim. 
Notice first that $AdS_5 \times S^5$ is a special case of the solution, 
eq.~(\ref{geomdil}).
When ${\tilde g}_{\mu\nu}$ and $\Phi$ are small perturbations about the 
$AdS_5 \times S^5$ solution,
\begin{subequations}
\begin{align}
\label{spaert}
{\tilde g}_{\mu\nu} & =\eta_{\mu\nu} + h_{\mu\nu}, \\
\Phi(x) & = \ln g_s + \delta \Phi(x),
\end{align}
\end{subequations}
we can use the standard AdS/CFT dictionary and it is easy to see that 
there is a dual field theory description for these backgrounds. Each 
supergravity perturbation has a normalisable and a non-normalisable mode 
\cite{vijay9805, vijay9808} associated with it; these respectively
determine the 
expectation value of the corresponding operator, and the source coupling 
to the operator in the dual theory.  For the solution, eq.~(\ref{geomdil}), 
we see that only the non-normalisable modes are
turned on in these backgrounds. Thus the dual theory in the linearised
perturbation case, is the ${\mathcal N}=4$ theory, in the ${\mathcal N}=4$
vacuum, with the additional source terms,
\be
\label{sources}
S_{source}=\int d^4x\ [\frac{\delta\Phi(x^\mu)}{g_\text{YM}^2} \Tr F^2
+h_{\mu\nu}T^{\mu\nu}].
\ee

For a background which is not a small perturbation about $AdS_5\times
S^5$ the above argument does not directly apply.  In this case it is
useful to look at things from the perspective of the gauge theory. We
are interested in the ${\mathcal N}=4$ gauge theory subjected to sources,
${\tilde g}_{\mu\nu}, e^\Phi$.  Suppose we are also interested in a
situation where the gauge theory is in the ${\mathcal N}=4$ vacuum in the
far past, as $X^+\rightarrow -\infty$. This is a reasonable initial
state, since the sources, ${\tilde g}_{\mu\nu}, \Phi$, become the
flat metric and a constant dilaton respectively, as $X^+\rightarrow
-\infty$.  The initial conditions and the sources specify the gauge
theory completely.  Now it is reasonable to believe that there is a
supergravity dual for this gauge theory. Since the data discussed
above specifies the gauge theory completely, the supergravity solution
should be unique.  The solutions we have given in section~\ref{sec:soln}
meet all
the required conditions.  Eqs.~(\ref{geom}) and~(\ref{formdil})
have the correct asymptotic behaviour
as $r\rightarrow \infty$ to turn on
the source terms ${\tilde g}_{\mu\nu}$ and $e^\Phi$; the
solutions~\eqref{geomdil}
also reduce to the $AdS_5\times S^5$ solution as $X^+\rightarrow
-\infty$.  Thus they must be the supergravity dual to the gauge
theory.

It is also worthwhile examining this issue from the bulk viewpoint. 
We can try to devise  an argument along the lines of the one used to 
motivate the AdS/CFT correspondence. 
One can show \cite{dmnt0602} that a background of the form, 
\be
ds^2 = Z^{-1/2}(x) \tilde{g}_{\mu\nu} dx^\mu dx^\nu
+ Z^{1/2}(x)  dx^m dx_m, \qquad \Phi=\Phi(x^{\mu}),
\ee
with the self-dual  five-form, $F=*_{10}F$, with nonzero components, 
\be
F_{0123m}=\frac{1}{4 \kappa}\frac{1}{Z \sqrt{det(-{\tilde g_{mn}})}}\ 
\partial_m\log Z,
\ee
and its dual
satisfies the equations of motion, as long as ${\tilde g}_{mn}$ and
$\Phi(x)$ satisfy the conditions eq.~(\ref{conds}), and $Z$ is a harmonic
function on the flat six-dimensional space with coordinates, $x^m,
m=1\cdots 6$.  The solution, eq.~(\ref{geom}), corresponds to taking,
$Z=R^2/r^2, r^2=x^mx_m$. Another choice is to take
\be
\label{fullh}
Z=1+\frac{R^2}{r^2} .
\ee
The solution, eq.~(\ref{geom}), can then be obtained by the 
``near-horizon limit'', $r\rightarrow 0$, from this case. 

In more detail, for the choice, eq.~(\ref{fullh}), the asymptotic form of 
the metric and dilaton, as $r\rightarrow \infty$, are,
\be
ds^2 =\tilde{g}_{\mu\nu} dx^\mu dx^\nu + dx^mdx_m, \qquad 
\Phi=\Phi(x^\mu).
\ee
One can verify that this background (without any five form flux)
solves the equations of motion.  Now we can add D3-branes to this
background. In the probe approximation it is easy to see that each
D3-brane will see a four dimensional metric along its world volume of
the form, ${\tilde g}_{\mu\nu}$, and a gauge coupling
$g_\text{YM}^2=e^\Phi$. Adding $N$ D3-branes and taking a low-energy limit,
as in the usual AdS/CFT case, would then lead to the field theory dual
for the solution, eq.~(\ref{geomdil}), mentioned at the
beginning of this section.  As an additional final check we note that a
probe D3 brane added to the background, eq.~(\ref{geomdil}),
also sees a four dimensional metric along its world
volume and a gauge coupling given by, ${\tilde g}_{\mu \nu}$,
$g_\text{YM}^2=e^{\Phi}$, respectively\footnote{The DBI action for a probe 
D3-brane gives
\begin{equation*}
\int d^4x\ e^{-\Phi} \sqrt{-\det (G^{str}_{ab} + F_{ab})}
= \int d^4x\ \sqrt{-\det ({\tilde g}_{\mu\nu})}\ \left(
\frac{1}{2}{\tilde g}^{\mu\nu}\del_{\mu}x^m \del_{\nu}x^m-\frac{1}{4}e^{-\Phi}
{\tilde g}^{\mu\rho}{\tilde g}^{\nu\sigma}F_{\mu\nu}F_{\rho\sigma}+\ldots
\right),
\end{equation*}
where $G^{str}_{\mu\nu}=e^{\Phi/2}{\tilde g}_{\mu\nu}$ is the string 
metric.}.

In the rest of the paper we will explore the consequences assuming
that this conjectured duality with the ${\mathcal N}=4$ field theory
is correct.  Before moving on, though, it is worth emphasizing that
the arguments given above in support of the dual field theory
description while plausible are not air-tight.  In particular, they
are not on as firm footing as for the original AdS/CFT correspondence
and might especially fail close to the singularity in the bulk, where
the low-energy limit is more subtle. An important check in the AdS/CFT
case was that the absorption cross-sections~\cite{cs1,cs2,cs3}
vanished at low energies in agreement with the required decoupling of
the near and far-horizon regions. We have not attempted to devise
analogous checks in the case at hand. It is probably useful for this
purpose to regard the singular solution, eq.~(\ref{nullsolntanh}) as
the limiting case of the one-parameter family, eq.~(\ref{onep}).  Such
checks might reveal subtleties close to the space-time singularity.

It is worth noting that in principle one could study the \Nf\ SYM gauge 
theory with more general time-dependent $e^\Phi, {\tilde g}_{\mu\nu}$, 
deformations, but identifying their supergravity duals might be 
difficult.

We end this section with one comment that will be important in the 
subsequent discussion. In the null metric, eq.~(\ref{metpar}), the 
conformal anomaly for the ${\mathcal N}=4$ theory vanishes. The conformal 
anomaly is given by \cite{duff20yrs, osborn9307,
anselmi9601, erdmenger9605, anselmi9708} (see also \eg\
\cite{skenderis9806, skenderis9812, balakraus9902, magoo9905} in the
holographic context of $AdS_5\times S^5$), 
\be
\label{ca}
T_{\mu}{^\mu} =
\frac{c}{16\pi^2}(C_{\alpha\beta\gamma\delta}C^{\alpha\beta\gamma\delta})
- \frac{a}{16\pi^2}(R_{\alpha\beta\gamma\delta}R^{\alpha\beta\gamma\delta}
- 4R_{\alpha\beta}R^{\alpha\beta} + R^2)\ \propto\ \
-R_{\alpha\beta}R^{\alpha\beta} + \frac{1}{3}R^2,
\ee
where in the last expression we have used $a=c=\frac{N^2-1}{4}$ for the
$SU(N)$ \Nf\ super Yang-Mills theory. 
In the null solutions eq.~(\ref{metpar}), since $R_{++}$ is the only
non-vanishing component of the stress tensor, the conformal anomaly
vanishes. 

The null solutions eq.~(\ref{metdil}), also have a
varying dilaton.  In general, this could give rise to an additional
contribution in the trace anomaly.  However, such a contribution can
be ruled out for these solutions by the following argument.  The
additional term must be generally covariant involving derivatives of
the dilaton, the metric and tensors made out of the metric, like the
Ricci curvature. But any such term evaluated on the solutions
eq.~(\ref{metdil}) vanishes because the
only derivative of the dilaton that is nonvanishing is
$\partial_+\Phi$, while the metric component ${\tilde g}^{++}$ and
similarly the component of any tensor made out of the metric with two
upper $+$ indices vanishes\footnote{We thank members of the TIFR
string theory group and especially Shiraz Minwalla for discussions in
this regards and for providing this argument.
A concrete realization of this argument in this context can be found
in~\cite{confdil1,confdil2}; see for example eq.~(25) of~\cite{confdil1}
and eq.~(24) of~\cite{confdil2}.
\label{ft:TIFR}
}.
 
\section{Analysis of the Singularity in the Dual Field Theory}
\label{sec:analysis}

In this section we analyse the ${\mathcal N}=4$ Super Yang-Mills theory in
a space-time with metric ${\tilde g}_{\mu \nu}$ and gauge coupling
$g_\text{YM}^2=e^\Phi$ given in eqs.~(\ref{geom}) and~(\ref{formdil}).  We are
interested in the case of conformally flat null backgrounds,
eq.~(\ref{metpar}), which asymptotically become $AdS_5\times S^5$, as
$X^+ \rightarrow -\infty$, and in which the dilaton remains weakly
coupled for all times. In the examples of interest a singularity
arises when the conformal factor $e^f \rightarrow 0$, and this happens
at finite affine time.  We will choose coordinates so that the
singularity occurs at $X^+=0$. A prototypical example is
eq.~(\ref{nullsolntanh}).

Our main conclusion is that appropriately defined correlation functions
in the gauge theory are nonsingular at the the singularity, \ie\ at
$X^+=0$.  Physically the singularity arises because the metric shrinks
to zero. The essential reason why the field theory is nonsingular is
that the metric is conformally flat and since the ${\mathcal N}=4 $ Super
Yang-Mills theory is conformally invariant, appropriately defined
correlation functions do not depend on the conformal factor, even when
it vanishes.  Now the Yang-Mills theory we are interested in also has
a varying gauge coupling. This leads to a non-trivial dependence of
correlators on the background,
but we will argue that the resulting theory is still
nonsingular.

The analysis is divided into two parts. In the next subsection we neglect
the variation of the gauge coupling and analyse the field theory in a
conformally flat background.  Here, our discussion is for a general
conformally invariant theory.  The following section then includes the
effects of the varying gauge coupling in the ${\mathcal N}=4$ Super Yang
Mills theory.

\subsection{Conformal Field Theory in a Conformally Flat Background}

Consider a conformally invariant field theory in a conformally flat 
background, 
\be
\label{cfmet}
ds^2=e^f \eta_{\mu\nu} dx^\mu dx^\nu,
\ee
and consider operators ${\mathcal O}_i$ with \cfd dimensions $\Delta_i$, 
in this theory.
Then it is straightforward to show that  correlation functions of these 
operators,  dressed by  powers of the 
conformal factor determined by their \cfd dimensions, in the background
eq.~(\ref{cfmet}), are the same as in flat space. That is, 
\be
\label{genindt}
\langle e^{\frac{f(x_1)\Delta_1}{2}} {\mathcal O}(x_1)
e^{\frac{f(x_2) \Delta_2}{2}}
{\mathcal O}(x_2) \cdots 
e^{\frac{f(x_n) \Delta_n}{2}} {\mathcal O}(x_n)\rangle_{[e^f \eta_{\mu\nu}]}=
\langle{\mathcal O}(x_1) {\mathcal O}(x_2) \cdots {\mathcal O}(x_n)
\rangle_{[\eta_{\mu\nu}]}.
\ee
We are using notation where the background metric is denoted within
square brackets suffixed to the correlator. Thus the left-hand side (lhs)
is evaluated
with the metric, $e^f\eta_{\mu\nu}$, while the rhs
is in flat space.
One important condition must be met by the metric, eq.~(\ref{cfmet}),
for the result, eq.~(\ref{genindt}), to be true. The trace of the
energy-momentum tensor, \ie\ the conformal anomaly $T^\mu{_\mu}$ in the
metric, eq.~(\ref{cfmet}), must vanish.

To establish eq.~(\ref{genindt}) we first start with the partition function 
of the conformal field theory in a general background metric, $g_{\mu\nu}$. 
We denote the fields in the theory over  which the path integral is defined 
schematically by $\varphi$, and write, 
\be
\label{fone}
Z[g_{\mu\nu}]=\int [D\varphi]_{[g_{\mu\nu}]} e^{iS[g_{\mu\nu},\varphi]}.
\ee
Here, $S$ is the action of the CFT which depends on the fields $\varphi$ and 
the metric $g_{\mu\nu}$, and we have explicitly indicated the dependence 
of the measure on the background metric. 
Under an infinitesimal conformal transformation,
\be
\label{fthree}
g_{\mu\nu}\rightarrow e^{\delta \psi} g_{\mu\nu},
\ee
the change in the partition function is proportional to the trace of the 
energy-momentum tensor, $T^\mu{_\mu}$:
\be
\label{ffour}
\delta \log Z=i\langle\int d^4x \sqrt{-g}\ T^\mu{_\mu}\delta \psi\rangle.
\ee

Let us be more explicit in how eq.~(\ref{ffour}) is derived. From
eq.~(\ref{fone}), 
\be
\label{ffive}
Z[e^{\delta \psi} g_{\mu\nu}]=\int [D\varphi]_{[e^{\delta \psi}
g_{\mu\nu}]}~ e^{iS[e^{\delta \psi}g_{\mu\nu},\varphi]}
\ee
If the fields $\varphi$ have \cfd dimensions $\Delta$, we now change 
variables in the path integral from $\varphi$ to ${\tilde \varphi}$ given by
\begin{equation}
{\tilde \varphi}=e^{\frac{\Delta \delta \psi}{2}} \varphi,
\end{equation}
and then write,
\be
\label{fsix}
Z[e^{\delta \psi} g_{\mu\nu}]=\int [D{\tilde \varphi}]_{[g_{\mu\nu}]} 
e^{iS[g_{\mu\nu}, {\tilde \varphi}]} (1+i\int d^4x\sqrt{-g}\ 
T^\mu{_\mu}\delta\psi).
\ee
In general the term proportional to $T^\mu{_\mu}$ on the rhs arises both due 
to the change in the action and the change in the measure. For a conformal 
field theory the contribution arises entirely due to the change in the 
measure. Noting that ${\tilde \varphi}$ is a dummy variable in the path 
integral  we can then subtract eq.~(\ref{fone}) from eq.~(\ref{fsix})
giving eq.~(\ref{ffour}).

Now consider a one-parameter family of metrics, 
\be
\label{fnine}
g_{\mu\nu}(x)=e^{\alpha f(x)}\eta_{\mu\nu},\qquad \alpha \in [0,1].
\ee
If  $T^\mu{_\mu}$ vanishes for all values of $\alpha \in [0,1]$, we learn 
from eq.~(\ref{ffour})
that $\partial_\alpha Z=0$, so that  
\begin{equation}
Z[e^f \eta_{\mu\nu}]=Z[\eta_{\mu\nu}].
\end{equation}

This argument can be easily extended to correlation functions. Consider 
the two-point function of the dressed fields, 
$e^{\frac{f(x) \Delta}{2}} \varphi(x)$, evaluated in the metric, 
eq.~(\ref{cfmet}),
\be
\label{tpt}
\langle e^{\frac{f(x) \Delta}{2}} \varphi(x)
e^{\frac{f(y) \Delta}{2}} \varphi(y)
\rangle_{[ e^f \eta_{\mu\nu}]}= 
\frac{1}{Z[e^f \eta_{\mu\nu}]} \int [D \varphi]_{[ e^f \eta_{\mu\nu}]}
 e^{iS[ e^f \eta_{\mu\nu},\varphi]} e^{\frac{f(x) \Delta}{2}}
e^{\frac{f(y) \Delta}{2}} \varphi(x) \varphi(y)
\ee
One can show that this correlator is the same as the two point function 
of $\varphi(x)$ in flat space. That is, 
\be
\label{indt}
\langle e^{\frac{f(x)\Delta}{2}} \varphi(x)
 e^{\frac{f(y) \Delta}{2}}\varphi(y)
\rangle_{[e^f \eta_{\mu\nu}]}=
\langle \varphi(x)\varphi(y)\rangle_{[\eta_{\mu\nu}]}
\ee

The idea is to again ``build up'' the metric, eq.~(\ref{cfmet}), starting 
from the flat one, by considering the one-parameter family, 
eq.~(\ref{fnine}), and increasing $\alpha$ from $0$ to unity. 
We start with 
\begin{multline}
\label{ctpt}
e^{\frac{(\alpha + \delta \alpha) f(x) \Delta}{2}}
e^{\frac{(\alpha + \delta \alpha) f(y) \Delta}{2}}
\langle\varphi(x)\varphi(y)\rangle_{[
e^{(\alpha + \delta \alpha) f}\eta_{\mu\nu}]}
\\ =
\frac{1}{Z} \int [D \varphi]_{[e^{(\alpha + \delta \alpha )f} \eta_{\mu\nu}]}
e^{iS[e^{(\alpha + \delta \alpha) f} \eta_{\mu\nu},\varphi]}
e^{\frac{(\alpha + \delta \alpha ) f(x) \Delta}{2}}
e^{\frac{(\alpha + \delta \alpha)  f(y) \Delta}{2}}
\varphi(x) \varphi(y).
\end{multline}
We will again assume that $T^\mu{_\mu}$ vanishes for the one parameter family 
of metrics, eq.~(\ref{fnine}).
Since we have already shown that the partition function $Z$ is independent 
of $\alpha$, any change in the two-point function
as $\alpha$ changes  can only arise from the path integral in the numerator. 
Now changing the conformal factor of the background metric on the rhs from 
$e^{(\alpha + \delta \alpha)f} $ to
$e^{\alpha f}$ and changing variables in the path integral to
$\tilde \varphi=e^{\frac{\delta \alpha f \Delta}{2}} \varphi$ gives,
\begin{equation} \label{threept}
\begin{split}
e^{\frac{(\alpha + \delta \alpha) f(x) \Delta}{2}}
e^{\frac{(\alpha + \delta \alpha)  f(y) \Delta}{2}}
\langle\varphi(x)\varphi(y)\rangle_{[e^{(\alpha + \delta \alpha) f} 
\eta_{\mu\nu}]} 
& =
\frac{\int D {\tilde \varphi}_{[e^{\alpha f} \eta_{\mu\nu}]} 
e^{iS[e^{\alpha f } 
\eta_{\mu\nu},{\tilde \varphi}]}
e^{\frac{\alpha f(x) \Delta}{2}}
e^{\frac{\alpha f(y)\Delta}{2}} {\tilde \varphi(x)}
{\tilde \varphi(y)}}{Z} 
\\
&= e^{\frac{\alpha  f(x) \Delta}{2}}
e^{\frac{\alpha  f(y) \Delta}{2}}
\langle\varphi(x)\varphi(y)\rangle_{[e^{\alpha  f} \eta_{\mu\nu}] }.
\raisetag{\baselineskip}
\end{split}
\end{equation}

The rhs of the line above arises by noting that $T^\mu{_\mu}$ vanishes
for the particular background metric under consideration. The rhs of
the second line arises from our definition of the two point function,
eq.~(\ref{tpt}), after noting that ${\tilde \varphi}$ is a dummy variable
in the path integral.  From eq.~(\ref{threept}), we see that the
two-point function of the dressed operator does not change under an
infinitesimal change in $\alpha$. It then follows that the two-point
function is independent of $\alpha$ leading to eq.~(\ref{indt}).

It is now clear that this argument generalises for $n$-point correlators
of any set of operators ${\mathcal O}_i$, leading to the result,
eq.~(\ref{genindt}).

We have worked in Minkowski space above. This is the relevant setting
in the present investigation where we have a time dependent or null
background. In Minkowski space we have to specify the state of the
system in which the correlator is being computed. The more precise
version of eq.~(\ref{genindt}) is that the correlator on the left hand
side is evaluated in the conformal vacuum appropriate to the
metric, eq.~(\ref{cfmet}), while the correlator on the right hand side
is evaluated in the Minkowski vacuum. Our definition of the conformal
vacuum is the standard one in the discussion of quantum field theory
in curved space \cite{BD}. To avoid ambiguities we will discuss 
eq.~(\ref{indt}) in more detail for the concrete case of a conformally
coupled scalar field in 4 dimensions in Appendix~\ref{sec:confscalar}.

The above discussion has been for a general conformal field theory. It
also applies to the ${\mathcal N}=4$ super Yang-Mills theory. The only
additional condition that needs to be met is that the trace anomaly
vanishes. This is true for all backgrounds of the type,
eq.~(\ref{metpar}), as was discussed in eq.~(\ref{ca}) and in the subsequent
paragraph\footnote{The one
parameter family, eq.~(\ref{fnine}), can be obtained by taking the
conformal factor in eq.~(\ref{metpar}) to be $e^{\alpha f(X^+)}$. Then
it is easy to see that the conformal anomaly vanishes for all values
of $\alpha$.}. 

It is also worth mentioning that in our discussion above, conformal
invariance really means Weyl invariance, \ie\ position dependent
rescalings of the metric without coordinate transformations. The
conformal dimension of an operator, $\Delta$, which appears in the
dressing factor above, is therefore determined by how the operator
transforms under a Weyl transformation.  Thus in the ${\mathcal N}=4$
theory, the gauge potential $A_\mu$, which does not transform under a
Weyl transformation, has $\Delta=0$, while, $F^2\equiv
F_{\mu\nu}F_{\rho\sigma}g^{\mu\rho} g^{\nu\sigma}$, has $\Delta=4$.  In
particular we see that $\Delta$ can be different from the dimension of
an operator under a rescaling of the coordinates. The gauge potential
$A_\mu$, for example, has dimension $1$ under such rescalings.

We are especially interested here in the case where the conformal
factor $e^f$ shrinks to zero resulting in a singularity. One might be
worried that our conclusions above fail at the singularity.  It is
useful to think of the singular case as the limiting situation in a
one-parameter family.  For the case, eq.~(\ref{nullsolntanh}), this
family is described in eq.~(\ref{onep}) and is labelled by the
parameter $\epsilon$.  For all values of $\epsilon>0$, our discussion
above does apply, since the conformal anomaly vanishes for all values
of the parameter $\epsilon$, and eq.~(\ref{genindt}) is valid\footnote{
Since the metric, eq.~(\ref{onep}), is   non-analytic at $X^+=0$, some   
derivatives of the metric  have a  finite discontinuity there.
However, for both $X^+ \rightarrow 0^+$ and $X^+ \rightarrow 0^-$,
the arguments  for the vanishing of the conformal anomaly apply.}.  
   We can
define the correlation functions in the theory with $\epsilon=0$ as
the limiting case obtained by taking $\epsilon\rightarrow 0$. We then
conclude that the dressed correlators in the case with the vanishing
conformal factor equal those in Minkowski space and in particular are
nonsingular\footnote{ The variation of the dilaton is not being
included here and will be analysed in the subsection below. Let us
note for now that in the family, eq.~(\ref{onep}), at $X^+=0$,
one has $e^\Phi \sim g_s (\epsilon)^{\sqrt{8}}$. So for small
$g_s$, the string coupling is non-vanishing but small.}.

\subsection{Effects of the Varying Dilaton}

We now turn to incorporating the effects of the varying dilaton in the
Super Yang-Mills theory.

The supergravity backgrounds we are interested in are asymptotically,
as $X^+\rightarrow -\infty$, $AdS_5 \times S^5$ with a constant
dilaton. For example, in eq.~(\ref{nullsolntanh}), as $X^+\rightarrow
-\infty$, we find that,
\be
\label{vardil}
\frac{ d e^{\Phi}/dX^+}{e^{\Phi}}=-4 \sqrt{2} e^{-|X^+|} \rightarrow 0.
\ee
Thus asymptotically the Yang-Mills theory is in a spacetime with flat
metric and constant dilaton.  Near the singularity, as $X^+\rightarrow
0$, $e^\Phi$ goes to zero and the Yang-Mills theory becomes free so
once again effects of the dilaton are not serious.  However, for
intermediate values, $X^+\sim O(1)$, the variation of the dilaton is
of order unity.  To obtain a well defined supergravity description in
the far past we need to take $g_sN\gg 1$. So in the intermediate
region, when $X^+\sim O(1)$, the Yang-Mills theory has a large and
varying 't Hooft coupling.

There are two concerns about such a varying coupling constant that we
address below.  First, this variation of the dilaton could 
potentially
also lead
to particle production.  If such particle production occurs, even if
we started in the vacuum of the ${\mathcal N}=4$
theory in the far past we would
not in general remain in the conformal vacuum at later times. For
example consider the situation when the big crunch at $X^+=0$ is
approached; our arguments above, which assumed that the theory was in 
the conformal vacuum, will now not apply.  Second, such $X^+$
variation might destroy the conformal invariance of the theory and
thus render our analysis of the previous section invalid.
  
It is useful to first consider a toy model of a conformally coupled
scalar field in $3+1$ dimensions subject to a null dependent
perturbation, $\int d^4x \sqrt{-g}J(X^+) \varphi^3$, to analyse both
issues.  We do so below and then return to the Super Yang-Mills case.

Our conclusions are that due to the variation of the dilaton being
null, \ie\ along a light-like direction, there is no particle
production\footnote{A more precise statement will be made
below.}. And while the variation of the dilaton destroys conformal
invariance, suitably defined correlation functions are expected to
be nonsingular in a background of the form, eq.~(\ref{metpar}),
eq.~(\ref{nullsolntanh}).

\subsubsection{The Conformally Coupled Scalar}

We consider a conformally coupled scalar with Lagrangian, 
\be
\label{lag2}
S=\int d^4x \sqrt{-g}\ [\frac{1}{2} (\partial\varphi)^2 
+ \frac{1}{6} R \varphi^2 + J(X^+) \varphi^3],
\ee
in a metric, 
\be
\label{metc}
ds^2=e^{f(X^+)} (-2dX^+dX^- + dx_i^2).
\ee
The light-cone quantisation of this theory without the $J(X^+)\varphi^3$ 
term is discussed in Appendix~\ref{sec:confscalar}.

\paragraph{Particle Production}

Let us now consider the effects of the perturbation, 
\be
\label{pert}
S_\text{pert}=\int d^4x \sqrt{-g} J(X^+) \varphi^3.
\ee

We take the operator $\varphi^3$ to be normal ordered with respect to the 
creation and anihilation operators, $a,a^\dagger$  defined in eq.~(\ref{mecf}).
 
In the interaction picture the resulting state of the system is given by,
\be
\label{intpic}
|s\rangle=T_+ e^{-i\int d^4x e^{2f(X^+)} J(X^+) \varphi^3} |0\rangle
\ee
The $T_+$ symbol refers to time ordering with respect to the $X^+$ direction. 
At first order we get,
\be
\label{fo}
\delta_1 |s\rangle=-i\int d^4x e^{2 f(X^+)} J(X^+) \varphi^3 |0\rangle
\ee

Now from eq.~(\ref{mecf}) we see that if $\varphi^3$ is normal ordered
only the $(a^\dagger)^3 $ term in $\varphi^3$ will survive. But each of
these $a^\dagger$ terms carries a positive momentum in the $X^-$
direction (momentum in the $X^-$ direction is the value of $k_-$, so
this means that each factor of $a^\dagger$ comes with a factor $e^{i
k_- X^-}$ where $k_->0$). Since $J$ is independent of $X^-$ the
integral over $dX^-$ (which has range $[-\infty,\infty]$) leads to a
delta function which means the sum of the three momenta along the
$k_-$ directions coming from each of the $a^\dagger$ terms must
vanish. Since each of these terms has a positive $k_-$ momentum
this constraint cannot be met and thus the first order term
eq.~(\ref{fo}) vanishes.

In fact it is easy to see that this argument generalises to all orders
in perturbation theory and applies even beyond perturbation theory.

To understand what is happening physically let us first consider an
analogous situation where the perturbation is not time dependent but
depends on a spatial coordinate. In this case, if we start with the
vacuum state we cannot produce any particles and the state must remain
in the vacuum. Producing particles means adding positive energy, but
time translational symmetry prevents that from happening.  Here, we
have a similar situation except since we are dealing with a null
perturbation it is a little less obvious. It is in fact the momentum
along the $X^-$ direction that plays the role of the energy. Starting
with the vacuum and adding particles means adding positive momentum
along the $X^-$ direction. But since the source term $J(X^+)$ does not
break translational invariance along $X^-$ this is not allowed and
thus the vacuum stays the vacuum.

The summary of the above analysis is that a source term which depends
on a null direction does not lead to particle production, and is more
analogous, as far as the question of particle production is concerned,
to a source which is spatially varying\footnote{Of course, there is
no invariant notion of a particle in general---it is observer and
vacuum dependent.  This also means that particle production is
observer dependent. In our analysis above the vacuum is the vacuum of
the free non-interacting theory and is kept unchanged.  And we then
find that in the interaction picture this state will not change during
time development, if the source term is null dependent. This is the
more qualified sense in which there is no particle production due to
a null dependent source.}.

Two comments are now in order. 

First, there can  be other situations of course  where there is particle 
production. Consider for example a free scalar field theory whose kinetic
term is
\ben
S = \int d^4x [F (\partial \varphi)^2]
\een
If $F$ is a function of time, rather than $X^+$, there would
be particle production in this model in the sense that the out vacuum
is related to the in vacuum by a nontrivial Bogoliubov transformation. 
In such a situation the out vacuum would be a squeezed state of in 
particles. However if $F$ is a function of $X^+$, arguments
identical to those given in the previous paragraphs ensure that there
is no particle production. The out vacuum, whatever it might be, must
contain a superposition of arbitrary number of 
{\em pairs} of in particles with a 
vanishing net $k_-$, and this cannot happen since there are no
particles with negative $k_-$.

Second, the above argument could have
subtleties for the $k_-=0$ modes. We have not examined this issue very
carefully: such complications are of course well-known in light-cone
quantisation (for example, see \cite{lc,lc2}).  Ultimately we use the fact
of no particle production to make some statements about correlators.
As long as the correlators are not at zero $k_-$ momentum, we do not
expect to be very sensitive to this subtlety.

The source, $J(X^+)$, does have effects of course; it affects
correlation functions in the ground state. We turn to examining these
next.

\paragraph{Correlation Functions}
The field $\varphi$ in the conformal theory, eq.~(\ref{lag2}), has \cfd 
dimension $1$.
We consider the dressed Feynman two-point function in the presence of the 
source $J(X^+)$. This is given by, 
\be
\label{fcor}
G_F(x_1,x_2) = \langle0|T_+ e^{\frac{f(x_1)}{2}} \varphi(x_1) 
e^{\frac{f(x_2)}{2}} 
\varphi(x_2) e^{-i\int d^4x \sqrt{-g} J(X^+) \varphi^3(X)}|0\rangle
\ee
where the time ordering refers to $X^+$ ordering.
Now we see that when $X^+$ lies in between $x_1^+$ and $x_2^+$ then terms 
proportional to the source term $J(X^+)$ will not vanish. And thus the 
Feynman correlation function will depend on $J(X^+)$.

Some comments are now worth making. 
First, in the example eq.~(\ref{fcor}), it follows from our discussion 
in the previous section that if $e^{\frac{f(X^+)}{2}} J(X^+)$ vanishes as 
$X^+ \rightarrow 0$ then correlation functions where both $x_1^+,x_2^+$ 
are close to  the origin will to good approximation be
the same as in the case when $J(X^+)$ vanishes identically. To see this, 
we write the interaction Hamiltonian in eq.~(\ref{fcor}), as,
\be
\label{intH}
\int d^4X\sqrt{-g} J(X^+) \varphi^3(X)=\int d^4X e^{\frac{f}{2}} J(X^+) 
e^{\frac{3f}{2}} \varphi^3(X)
\ee
Now, we know that the dressed operator $e^{\frac{3f}{2}} \varphi^3(x)$ has 
nonsingular correlation functions, in the background eq.~(\ref{metc}) 
(since the \cfd dimension of $\varphi^3(x)$ is $3$).
Thus if $e^{\frac{f(X^+)}{2}}J(X^+)$ vanishes, as $X^+\rightarrow 0$,
the interaction Hamiltonian becomes negligible.

Second, in perturbation theory in $J(X^+)$, the correlation functions
with the source term can be related to those in the theory without the
source term. And the latter, we have seen in the previous section can
be related after appropriate dressing to correlation functions of the
CFT in flat space.  For example, to first order we see that
\begin{equation} \label{ffcor}
\begin{split}
G_F(x_1,x_2) &= -i \langle 0|T_+ e^{\frac{f(x^+_1)}{2}} \varphi(x_1) 
e^{\frac{f(x^+_2)}{2}} \varphi(x_2)\int d^4X \sqrt{-g} J(X^+) \varphi(X)^3 
|0\rangle\\
&= -i \int d^4X J(X^+) e^{\frac{f(X^+)}{2}}\langle 0|T_+ 
e^{\frac{f(x^+_1)}{2}} 
\varphi(x_1) e^{\frac{f(x^+_2)}{2}} \varphi(x_2)e^{\frac{3f(X^+)}{2}} 
\varphi^3(X) |0\rangle, 
\end{split}
\end{equation}
where in the second equation on the rhs we have dressed all the
operators by powers of $e^{f}$ proportional to their \cfd
dimensions.  Thus, in perturbation theory possible singular behaviour
could arise as $X^+\rightarrow 0$ only if $J(X^+) e^{f(X^+)/2}$
diverges as $X^+\rightarrow 0$.  The dressing factor, $e^{f/2}$, for
the source can be understood easily as it is simply determined by the
\cfd dimension of the operator $\varphi^3$. More generally for a
source $J(X^+)$ coupling to an operator of \cfd dimension
$\Delta$ the dressing factor would be $e^{\frac{4-\Delta}{2} f}$. As
long as $Je^{\frac{4-\Delta}{2}f}$ is well behaved as $X^+\rightarrow
0$, no singular behaviour will arise.

In the SYM theory which we turn to next, the source term is the dilaton
which couples to a dimension four operator. Since $e^\Phi$ vanishes
at the singularity no singular behaviour is expected in the correlation
functions.  The argument above was in perturbation theory.  In the SYM
theory perturbation theory is not a good approximation when the
't~Hooft coupling, $e^\Phi N\ge 1$.  However any singular behaviour is
expected to arise only at the singularity.  Since $e^\Phi$ vanishes
there, eq.~(\ref{nullsolntanh}), we expect that the conclusion that the
correlation functions are nonsingular should be more generally true.

We turn to a more detailed discussion of the SYM theory next.

\subsubsection{The SYM theory in the presence of a varying Dilaton}
\label{sec:varyingdil}

We now show that theories in such null backgrounds admit descriptions 
in terms of new tilde variables. Before discussing the gauge theory, we
first discuss the basic point in the context of a scalar $\varphi$ with 
kinetic term containing a nontrivial null-dependent factor $e^{-\Phi(X^+)}$.
\be
S = -\int d^4x\ e^{-\Phi(X^+)} (\del\varphi)^2 
\ee
If the function $e^{\Phi(X^+)}$ vanishes at some point, say $X^+=0$,
the propagator for $\varphi$ would vanish there as well and it might
appear that the theory is singular.
However we can define new
variables $\varphi=\epsilon(x){\tilde\varphi}$, so that 
the action becomes
\be\label{redefScalar}
S = -\int d^4x\ e^{-\Phi}\ \eta^{\mu\nu} \left(
\epsilon^2\del_{\mu}{\tilde\varphi}\del_{\nu}{\tilde\varphi} + 
\epsilon \del_{\mu}\epsilon \del_{\nu} ({\tilde\varphi}^2) + 
(\del_{\mu}\epsilon \del_{\nu}\epsilon) {\tilde\varphi}^2\right).
\ee
Choosing $\epsilon(x)=e^{\Phi(X^+)/2}$, we see that the third term, akin 
to a ``mass-term'', vanishes. Also given the null dependence of 
$\epsilon(x)$, the second (cross-)term becomes a total derivative\ 
$\del_+(\epsilon^2) \del_-({\tilde\varphi}^2)
= \del_- [\del_+(\epsilon^2) {\tilde\varphi}^2]$, which can be dropped.
Thus we see that such a theory with a null-dependent kinetic term can 
be described in terms of new variables ${\tilde\varphi}$ which have 
canonical kinetic terms. The essential point is that the Fock space
of this theory is defined in terms of creation and annihilation 
operators coming from the usual mode expansion of ${\tilde{\varphi}}$.

If we had started instead  with an interacting theory with action,
\be
\label{actint}
S=-\int d^4x \ e^{-\Phi(X^+)} [ (\del\varphi)^2 -\lambda \varphi^4],
\ee
then after changing to $\tilde \varphi$ variables and dropping a surface term,
the resulting action is,
\be
\label{actfin}
S=-\int d^4x[(\del {\tilde \varphi})^2 - \lambda e^{\Phi(X^+)}  {\tilde \varphi}^4].
\ee
We see that the ${\tilde \varphi}^4$ coupling is $X^+$ dependent. For a dilaton, $e^\Phi$,
 as in eq.~(\ref{nullsolntanh}), which is bounded and vanishing at 
$X^+ \rightarrow 0$, we see that the interaction term is also bounded and vanishing
at $X^+=0$. It is clear that this theory has a 
non-singular  S-matrix in perturbation theory and is therefore well-defined.
This is transparent in terms of the ${\tilde{\varphi}}$ variables which, as we noted,
are the ones relevant for defining asymptotic states.
 
In what follows, we will similarly find new 
nonsingular variables in our case of the SYM theory with null-varying 
dilaton.

The gauge coupling in the SYM theory is given by $g_\text{YM}^2=e^\Phi$. Thus 
a varying dilaton results in a varying gauge coupling. In more detail the 
dilaton couples to the terms in the
Lagrangian which are quadratic in the gauge field strength:
\be
\label{coudila}
S_\text{GF}
=-\frac{1}{4}\int d^4x\ \frac{1}{e^\Phi}\ \Tr[F_{\mu\nu} F^{\mu\nu}].
\ee
For now we will work with a flat metric. The effects of the 
nontrivial conformal factor $f(X^+)$ will be included in the discussion 
a little later.
Since the $e^\Phi$ vanishes at  the singularity, $X^+=0$,
the quadratic terms for  the gauge field $A_\mu$ become singular there. 
It is therefore useful to carry out a field redefinition to 
new variables which have well behaved quadratic terms. 
For simplicity we work in the gauge\footnote{The condition, $A_-=0$, 
leaves a residual freedom to do $X^-$ independent gauge transformations. 
This only affects modes with momentum $k_-=0$.
As was mentioned before, there are well-known subtleties associated 
with quantising the $k_-=0$ modes in light-cone gauge quantisation 
\cite{lc,lc2}, and we mainly consider modes with $k_-\ne 0$ in our 
analysis. For such modes the condition, $A_-=0$ fixes the gauge
completely.\label{ft:k-}},
\be
\label{gaugea}
A_-=0.
\ee
Then define 
\be
\label{defa}
{\tilde A}_i=e^{-\Phi/2}A_i, \qquad {\tilde A}_+=e^{-\Phi/2} A_+,
\ee
where the index $i$ in the first equation above takes values in the 
directions transverse to $X^+,X^-$. This gives 
\begin{multline}
\label{sgfbi}
S_\text{GF} = -\frac{1}{4}\int d^4x\ [\ \Tr(\partial_\mu {\tilde A}_\nu - 
\partial_\nu {\tilde A}_\mu)^2 
-2 i e^{\Phi/2} 
\Tr\{(\partial_\mu {\tilde A}_\nu - \partial_\nu {\tilde A}_\mu)
[{\tilde A}^\mu, {\tilde A}^\nu]\} \\
- e^\Phi \Tr([\tilde A_\mu,\tilde A_\nu])^2 - 
\partial_{X^-}\{(\partial_+\Phi){\tilde A}_i {\tilde A}^i\}\ ]
\end{multline}
In the last term the index $i$ takes two values transverse to both $X^+,X^-$. 
We see that the last term is a total derivative in $X^-$; it will 
not affect the equations of motion and can be dropped\footnote{The 
null-dependence of the dilaton was crucial for this to happen. If the 
dilaton were time-dependent instead, there would be additional terms in 
eq.~(\ref{sgfbi}), after the redefinition, eq.~(\ref{defa}), which are not 
total derivatives.}. The dilaton also appears in other couplings in the 
Lagrangian when we work in terms of the ${\tilde A}$ fields. These are
\be
\label{addcoupldil}
S=\int d^4x [ e^{\frac{\Phi}{2}} J^{\mu a} {\tilde A}_{\mu a}
+ e^{\Phi} \Tr([{\tilde A}_\mu ,\phi^\alpha][{\tilde A}^\mu,\phi^\alpha]) 
+ e^\Phi \Tr([\phi^\alpha ,\phi^\beta][\phi^\alpha, \phi^\beta])]
\ee
where $J^{\mu a}$ is the $SU(N)$ gauge current arising from the scalars 
and the fermions, and $\phi^\alpha, \alpha=1, \cdots 6$ are the six scalars. 

Before proceeding let us make two comments. 
First, in the gauge $A_-=0$, the 
variable, $A_+$ and therefore also ${\tilde A}_+$ is non-dynamical. 
The equation of motion for $A_-$ needs to be imposed as a constraint,
and this determines $A_+$ in terms of the transverse components $A_i$. 
Similarly, ${\tilde A}_+$ can be determined in terms of ${\tilde A}_i$.
The action in eq.~(\ref{sgfbi}) contains both ${\tilde A}_+$ and 
${\tilde A}_i$. The correct equations of motion for ${\tilde A}_i$ can be 
obtained from it by treating all of them  as independent variables, 
deriving the equations of motion for ${\tilde A}_i$ and then substituting 
for ${\tilde A}_+$ in terms of ${\tilde A}_i$ in them. Alternatively, the 
same equations of motion for ${\tilde A}_i$ can also be obtained by first 
substituting in the action, eq.~(\ref{sgfbi}), for ${\tilde A}_+$ in terms 
of ${\tilde A}_i$ and then varying with respect to ${\tilde A}_i$. 
Second, the ${\tilde A}$ variables can be also be  defined in terms of the 
 the gauge potential,
$A_\mu$, in a general gauge, as,
\be
\label{tildegen}
{\tilde A}_\mu=e^{-\Phi}(A_\mu+\partial_\mu \chi).
\ee
Where, $\chi$ is given by, 
\be
\label{defchi}
\chi=-\partial_-^{-1}A_-.
\ee
Note that eq.~(\ref{defchi}) uniquely defines $\chi$ as long as the momentum component,
$k_-$ is nonvanishing (subtleties related to the $k_-=0$ mode were discussed
in footnote~\ref{ft:k-}).

We see that in the terms remaining in eq.~(\ref{sgfbi}) after dropping
the total derivative in $X^-$, and in all the terms in
eq.~(\ref{addcoupldil}), $e^\Phi$ couples with positive powers. Thus
these couplings all vanish when $e^\Phi$ vanishes\footnote{In the
example, eq.~(\ref{nullsolntanh}), $e^{\Phi}$ is not an
analytic function of $X^+$ at the singularity and sufficiently high
order derivatives with respect to $X^+$ diverge. However the couplings
in the YM theory only involve positive powers of the dilaton and not
its derivatives. Thus these couplings all vanish at the singularity, 
and the diverging higher $X^+$-derivatives of the dilaton do not lead 
to any singular behaviour of the correlation functions.}. In fact these 
couplings are similar to the source terms considered in the discussion 
above of the conformally coupled scalar field. Therefore the
conclusions we reached in the conformally coupled case can also be
applied to the dilaton coupling in the SYM theory.  First, the
light-like variation of the dilaton will not give rise to any particle
production and will leave the ${\mathcal N}=4$ vacuum unchanged. Second,
it will have effects on the correlation functions. So far, we have not
included the nontrivial conformal factor in the metric. Once this is
included, it is important to work with appropriately dressed operators, 
as discussed in the previous section. Since the dilaton is a source
coupling to dimension four operators, we need to examine its behaviour
without any dressing factor of $e^f$. And since $e^\Phi$,
eq.~(\ref{nullsolntanh}), vanishes 
at the singularity $X^+=0$, we see that no 
singularities will arise in the dressed correlation functions\footnote{%
It is important to note that the varying dilaton does not change the
conformal dimensions of operators, assuming that any such contribution
to the
conformal dimension must come from a term in the effective action.
Since the effective action is a local coordinate invariant function of the
dilaton, the metric and the tensors made from the metric, any additional
terms
vanish  since the background is null. A similar argument was
discussed in the case of the trace anomaly at the end of
section~\ref{sec:dual}.}.
 For 
instance, the scalar kinetic and quartic interaction terms have factors 
of $e^f$ and $e^{2f}$ respectively, so that redefining the scalars to 
have canonical kinetic terms removes the $e^f$ dressing factors from 
both these dimension four operators. 

It is worth emphasizing that our arguments go through for the dressed
correlation functions constructed out of the ${\tilde A}_\mu$ fields
and their field strengths. As was mentioned above it is these fields
which have well defined quadratic terms at the singularity, where $e^\Phi$
vanishes. Other fields which are related by singular field
redefinitions to ${\tilde A}$ will not have nonsingular correlators
at the singularity in general.  For example, the field redefinition
used to go from the ${\tilde A}$ variables to the $A$ variables is
singular when $e^\Phi$ vanishes, eq.~(\ref{defa}).  If we had used the
field strength made out of the original $A_\mu$ variables,
eq.~(\ref{coudila}), then the correlation functions would not be finite
near the singularity.  Keeping only the quadratic terms in the field
strength and using the relation, eq.~(\ref{defa}), we have that
\be
\label{untilde}
e^{-\Phi}\Tr F^2 = \Tr \tF^2 - \frac{1}{2}\partial_-G_+,
\ee
where\ $G_+ \equiv (\partial_+\Phi)\tA_i \tA^i$. This means
\begin{equation}\label{gaugecorrelator}
\begin{split}
\langle e^{-\Phi}\Tr F^2(x) e^{-\Phi}\Tr F^2(y) \rangle
= \vphantom{a}& \langle \Tr \tF^2(x) \Tr \tF^2(y) \rangle \\
& - \frac{1}{2} \del_{x^-} \langle G_+(x) \Tr \tF^2(y)\rangle
- \frac{1}{2} \del_{y^-} \langle G_+(y) \Tr \tF^2(x)\rangle \\
& + \frac{1}{4} \del_{y^-} \del_{x^-}\langle G_+(y) G_+(x)\rangle
\end{split}
\end{equation}
Now for the  solution, eq.~(\ref{nullsolntanh}), close to the singularity,
$\Phi \sim \sqrt{8} \log X^+$. Thus $\partial _+\Phi \sim 1/X^+$, and 
diverges near the singularity. This means if one or both points in the 
correlator approach the singularity the $G_+$ dependent terms 
in eq.~(\ref{gaugecorrelator}) blows up. Thus the  correlation function of 
$e^{-\Phi}F^2$ diverges at the singularity\footnote{We have not 
explicitly put in any dressing factors proportional to $e^f$ in the 
correlation function, eq.~(\ref{untilde}). We are actually interested in 
the operator $\sqrt{-g} F_{\mu\nu} F_{\rho\sigma} g^{\mu\rho} g^{\nu
\sigma}$. For the conformally flat metric, eq.~(\ref{nullsolntanh})
factors of $e^f$ drop out. In the expression, $F^2$, in
eq.~(\ref{untilde}), the indices are contracted using the flat space
metric; similarly for ${\tilde F}^2$.}. 
This observation will be relevant for the discussion in the next
section where we compare the gauge theory results with those in
supergravity. The dilaton in the bulk couples to the operator,
$e^{-\Phi}F^2$, and thus the bulk two-point function for the dilaton
should be compared with the two-point correlator of this operator in
the SYM theory.

Before proceeding let us also note that the two-point function,
$\langle \Tr F^2(x) \Tr F^2(y) \rangle $, and all higher point functions
of $\Tr F^2$, vanish at the singularity for the example, eq.~(\ref{nullsolntanh}),
since $e^\Phi$ vanishes more rapidly at the singularity than $\partial_-G_+$.
More generally, depending on  the behaviour of  $e^\Phi$,
 these correlators could diverge at the singularity. 
  
It would be worthwhile to try and identify supergravity duals to the 
operators made out of the tilde variables. 
Some operators made out of ${\tilde A}$ fields can be easily related to local
operators made out of the  field strength,
$F_{\mu\nu} \equiv \partial_\mu A_\nu -\partial_\nu A_\mu -i[A_\mu, A_\nu]$.
For example, let us define,
\begin{equation}
{\tilde F}_{\mu\nu}=\partial_\mu {\tilde A}_\nu -\partial_\nu {\tilde A}_\mu
-ie^{\Phi/2} [{\tilde A}_\mu, {\tilde  A}_\nu].
\end{equation}
Then as long as $\mu,\nu\ne X^+$, we have a local  relation,
\be
\label{twot}
{\tilde F}_{\mu\nu}=e^{-\Phi/2} F_{\mu\nu}.
\ee
Some gauge invariant operators can be built from $F_{\mu\nu},\ 
\mu,\nu\neq X^+$,
by considering quadratic and higher powers and taking a trace over colour 
space. One example is $\Tr(F_{\mu\nu} F_{\rho\sigma})$. Such operators are 
dual to supergravity modes which are local excitations in the bulk.
Using eq.~(\ref{twot}) these can be described in terms of the tilde variables.
So for these operators, made out of the tilde variables, there is a dual 
description in terms of local excitations in the bulk.

But there are also operators, of a second type, made out of the tilde 
variables, which are not local functions of the field strength 
$F_{\mu\nu}$. The supergravity duals of these are not local excitations in 
the bulk, since the supergravity modes which are local excitations in the
bulk, are dual to gauge invariant operators made out of the field strength 
$F_{\mu\nu}$ and its powers. Example of such operators are ${\tilde A}_\mu$, 
or $(\partial_\mu {\tilde A}_\nu - \partial_\nu {\tilde A}_\mu)$.
Since the tilde variables are defined in the $A_-=0$ gauge, eq.~(\ref{defa}), 
these are by definition gauge invariant. 
But since the components of the gauge potential, $A_\mu$ cannot be 
expressed locally in terms of the field strength, neither can the fields 
${\tilde A}_\mu$. As another example,  take  ${\tilde F}_{\mu\nu}$, when 
either index, $\mu$ or $\nu$ is $X^+$. We get,
\be
\label{three}
{\tilde F}_{\mu\nu}= e^{-\Phi/2} F_{\mu\nu}
+ \frac{1}{2} e^{-\Phi/2} (A_\mu\partial_\nu\Phi-A_\nu\partial_\mu\Phi).
\ee
We see that the extra terms appearing on the rhs involve derivatives of 
$\Phi$, and the gauge potential $A$.
A complete set of operators made out of the tilde variables must include
operators of this second type whose supergravity duals are not local 
excitations in the bulk. 
 
It might be useful here to recall that a familiar example of an
operator which is gauge invariant but which is not local in terms of
the field strength is the Wilson loop, $\Tr Pe^{i\int A}$.  We know
this is dual to a string in the bulk and this is not a local excitation.
In some very rough sense, the duals of these second type of operators
should be similar.

In summary, in the last three sections we have analysed the dual gauge
theory in some detail.  Our conclusion is that nonsingular field
variables (the $\tilde A$ variables) can be found such that
correlation functions of these fields, when suitably dressed by
appropriate powers of the conformal factor, are nonsingular.  This
leads to the conclusion that the gauge theory dual is nonsingular.

\section{The Bulk Two-Point Function} \label{sec:green}

So far our discussion has been mainly in the gauge theory.  In this
section we calculate the two-point function of a scalar in the bulk.
Some comparisons with the gauge theory are discussed at the end of the
section.

It is convenient to work in coordinates in which the  $5$-dimensional 
space-time transverse to the $S^5$ in eq.~(\ref{geom}) has metric,
\be
\label{5dimmetric}
ds^2=\frac{1}{z^2} \Bigl[e^f \bigl(-2dX^+dX^-+\sum_{i=1,2} (dx^i)^2\bigr)
   + dz^2\Bigr].
\ee
The $5$ dimensional spacetime, eq.~(\ref{5dimmetric}) has a boundary at 
$z=\epsilon$.
This serves as an infrared regulator of the bulk theory. The dual gauge 
theory lives on the boundary as in the standard AdS/CFT correspondence.

A scalar of mass $m$ has the action 
\be\label{scalaraction}
S = -\int d^5x \sqrt{-g}\ (g^{\mu\nu}\del_{\mu}\varphi\del_{\nu}\varphi
+m^2\varphi^2),
\ee
A orthonormal complete set of modes solving the resulting wave equation is 
given by
\be
\label{finalsol}
u_{(k_i,k_-, \omega^2)}(z, x^\mu)=e^{-f(X^+)/2}\ 
e^{i(k_i^2X^+-\omega^2\int e^f dX^+)/2k_-} e^{ik_-X^-+ik_ix^i} \zeta_\omega,
\ee
where
\begin{equation}
\begin{aligned}
\zeta_\omega & = 
A \omega^2 z^2 K_{\nu}(\omega z) + B \omega^2 z^2 I_\nu(\omega z),
& \omega^2 &> 0, \\
\zeta_\omega & = A \omega^2 z^2 H_\nu(i \omega z) + 
B \omega^2 z^2 H_{\nu}(-i\omega z), & \quad \omega^2 &< 0.
\label{besselsol}
\end{aligned}
\end{equation}
Here $\omega \equiv \sqrt{\omega^2},\nu=\sqrt{4+m^2}$,
and $A,B$ are integration constants.

For the $AdS_5\times S^5$ background, if we are interested in calculating 
the bulk two-point function dual to the Feynman propagator in the 
${\mathcal N}=4$ vacuum of the gauge theory, then, for $\omega^2>0$, the 
correct combination of the normalisable and non-normalisable modes is 
obtained by setting $B=0$ in the first equation in (\ref{besselsol}), 
and only keeping the $K_{\nu}(\omega z)$ 
solution,
\be
\label{solc}
\zeta_\omega= (\omega z)^2 K_\nu(\omega z).
\ee
For $\omega^2<0$ the correct combination is obtained by analytically 
continuing this solution, 
\be
\label{sold}
\zeta_\omega =A \omega^2 z^2 H_\nu(i \omega z),
\ee
with $|A| = \frac{\pi}{2}$.
(Note the analytic continuation is not unique; one could have instead obtained 
\hbox{$\zeta = A\omega^2 z^2 H_\nu(-i\omega z)$}.
This ambiguity does not affect 
the final answer for the bulk two point function.)

The choice of these
modes is further justified by a standard light front quantization of the
field, which is explained in Appendix~\ref{sec:twopoint}.

We are interested here in backgrounds like eq.~(\ref{nullsolntanh}),
which become $AdS_5\times S^5$
asymptotically as $X^+\rightarrow -\infty$.
We have argued above that the dual SYM theory starts in the
${\mathcal N}=4$ vacuum in the far past, as $X^+\rightarrow -\infty$. To
calculate the bulk two-point function which is dual to the Feynman
propagator of the gauge theory in this case, we once again choose the
same combinations of normalisable and non-normalisable modes, as in
the $AdS_5 \times S^5$ case. This is clearly correct for correlation
functions in the limit, $X^+\rightarrow -\infty$, and since the $z$
dependent part is independent of $X^+$ it is must then true for all
$X^+$. 

A general solution can be expressed in terms of these modes,
\be
\label{modeexp}
\varphi(x^\mu,z) = \epsilon^{\Delta_-} \int_{-\infty}^{\infty}d^2k_i
\int_{-\infty}^{\infty}dk_-\int_{-\infty}^{\infty}\frac{d\omega^2}{2\abs{k_-}}\
\varphi(k_i,k_-, \omega^2) \ \frac{u_{(k_i,k_-, \omega^2)}(z,x^\mu)}{ 
\zeta_{\omega}(\epsilon)}.
\ee
$\varphi(k_i,k_-, \omega^2)$ are Fourier coefficients.   
$z=\epsilon$ is the boundary of space-time, as discussed after 
eq.~(\ref{5dimmetric}). We denote $\Delta_{\pm}=2\pm \nu$. 

The action evaluated on this classical solution becomes
\be
\label{simpli}
S=\int d^2k_i dk_- dk_+ C(\nu) \varphi(k_i,k_-, \omega^2) \varphi(-k_i,-k_-, 
\omega^2) \omega^{2\nu}
\ee
where the integrals  over all four variables, $k_i, i=1,2, k_-, k_+$ go 
from $[-\infty,\infty]$,
and $\omega^2=-2k_-k_+ + k_i^2$. 
An assumption  made in deriving  eq.~(\ref{valscalaract}) is that 
the affine parameter, $\lambda$ of eq.~(\ref{deflambda}),
has range $\lambda\in [-\infty,\infty]$.  This will be true if
the conformal factor obeys $e^f\ge 0$ for $X^+\in[-\infty,\infty]$,
since $d\lambda/dX^+=e^f \ge 0$. In position space, eq.~\eqref{simpli}
evaluates to 
\be\label{posnspAction}
S =C \int d^4x d^4x'\ e^{3f(X^+)/2} e^{3f({X'}^+)/2}\
\varphi({\vec x}) \varphi({\vec x'})\
\biggl(\frac{\Delta\lambda}{\Delta X^+}\biggr)^{1-\Delta}\ \frac{1}{
[(\Delta{\vec x})^2]^{\Delta}}.
\ee
Here, $C$ is a constant, and 
\be
\label{defD}
\Delta=2+\nu.
\ee

Let us relate this to a correlation function in the boundary theory. 
The mode $\varphi$ couples to an operator $O$  in the boundary theory with 
coupling,
\be
S_\text{Boundary}=\int d^4x \sqrt{-\tilde{g}}\ O(x) \varphi(x),
\ee
Here, $\tilde{g}_{\mu\nu} =e^f\eta_{\mu\nu} $, 
is the metric on the boundary and $\varphi(x)$ is given by eq.~(\ref{defp}).

Equating the bulk action with the  action of the boundary theory up to 
second order in the source $\varphi(x)$  gives, 
\begin{multline}
\sqrt{-\tilde{g}(x)} \sqrt{-\tilde{g}(x')} \langle O(x) O(x') \rangle
= \frac{\delta}{\delta\varphi({\vec x})} 
\frac{\delta}{\delta\varphi({\vec x'})}
\langle e^{\int d^4x\sqrt{-\tilde{g}} O(x)\varphi(x)} \rangle_\text{CFT}
\\
= \frac{\delta}{\delta\varphi({\vec x})} 
\frac{\delta}{\delta\varphi({\vec x'})}
e^{-S_\text{Sugra}[\varphi({\vec x})]}.
\end{multline}
From, eq.~(\ref{posnspAction}), we then get,
\be
\langle O(x) O(x') \rangle = C e^{-f(x)/2} e^{-f(x')/2} 
\biggl(\frac{\Delta\lambda}{\Delta X^+}\biggr)^{1-\Delta}\ \frac{1}{
[(\Delta{\vec x})^2]^{\Delta}}.
\ee

In the discussion on correlators of the gauge theory in
section~\ref{sec:analysis}
we
emphasized the importance of considering dressed correlators. The
operator $O(x)$ has \cfd dimension $\Delta$ in the SYM
theory. Then the dressed correlation function is
\be
\langle e^{\frac{f(x)\Delta}{2}} O(x) e^{\frac{f(x')\Delta}{2}} O(x')\rangle
= C e^{\frac{f(x)(\Delta-1)}{2}} e^{\frac{f(x')(\Delta -1)}{2}}
\biggl(\frac{\Delta\lambda}{\Delta X^+}\biggr)^{1-\Delta}\ 
\frac{1}{[(\Delta{\vec x})^2]^{\Delta}}  .
\ee

The behaviour of this correlator close to the singularity is worth analysing. 
Using the definition of the affine parameter $\lambda$, eq.~(\ref{deflambda}), 
we see that when the two points are close to each other,
\be
\label{twoclose} 
\langle e^{\frac{f(x)\Delta}{2}} O(x) e^{\frac{f(x')\Delta}{2}} O(x')\rangle 
=C \frac{1}{[(\Delta{\vec x})^2]^{\Delta}},
\ee
which is the two point function in the ${\mathcal N}=4$ SYM theory
without any sources. This is true when the two points are close
to each other in general and in particular when they are also close to
the singularity.  When one of the points, $\vec{x}$, is at the
singularity, $X^+=0$, we see that the dressing factor $e^f$ leads to
the correlation function vanishing. This is true for all values of
$(X')^+ \ne 0$. In the limit when both points approach the
singularity, $X^+,(X')^+ \rightarrow 0$, the correlation function
depends on how the limit is taken. For instance, for our prototypical 
example, eq.~(\ref{nullsolntanh}), we get, upon using $\lambda \sim (X^+)^3$
as $X^+\rightarrow 0$,
\be
\langle e^{\frac{f(x)\Delta}{2}} O(x) e^{\frac{f(x')\Delta}{2}} O(x')\rangle
= C\left(\frac{X^+}{(X')^+}\right)^{\Delta -1} \left[
\left(\frac{X^+}{(X')^+}\right)^2+\frac{X^+}{(X')^+}+1\right]^{1-\Delta} 
\frac{1}{[(\Delta{\vec x})^2]^{\Delta}}.
\ee 
The correlation function scales with distance $(\Delta{\vec x})^2
=(\Delta x_i)^2$ as in the ${\mathcal N}=4$ theory, but the coefficient 
depends on the limiting vaue of the ratio, $\frac{X^+}{(X')^+}$ . 

We argued in section~\ref{sec:varyingdil}
that close to the singularity the two point
function in the gauge theory reduces to that in free SYM theory, since
the dilaton vanishes.  We now see  that the
correlation function calculated from the bulk does not have this
property. The fall off with distance, $\frac{1}{[(\Delta{\vec
x})^2]^{\Delta}}$, is as in the free theory\footnote{ Bulk modes
correspond to operators whose \cfd dimensions are unrenormalised
and thus remain the same in the free limit.}, but as seen above,
 as the two points approach the singularity, the value of the
correlation function depends on the how the limit is approached and is
not unique.

This disagreement between the supergravity calculation and the gauge
theory does not disprove that the descriptions are dual.  The bulk
calculation fails close to the singularity where the higher derivative
corrections become important. Once these are incorporated presumably
the bulk answer will agree with the gauge theory.

Finally, the discussion above in particular applies to the dilaton,
which is a massless scalar in the five-dimensional theory. The operator it
couples to in the gauge theory is $O= e^{-\Phi} \Tr F^2$.  We discussed
in section~\ref{sec:varyingdil}
near eq.~(\ref{untilde}) that from the gauge theory point
of view the two point function of this operator should be singular
when one or both points approach the singularity.  This is very
different from the bulk result, which shows a result that is finite
but limit dependent. Once again presumably higher derivative
corrections are responsible for this disagreement.  The gauge theory
analysis also tells us that good variables in the gauge theory are the
${\tilde A}$ variables, eq.~(\ref{defa}), and gauge invariant field
strengths constructed out of these variables.  It would be interesting
to try and carry out a similar analysis for  bulk modes dual to these 
variables.

\section{Worldsheet Actions} \label{worldsheet}

In this section we consider the bosonic part of the worldsheet action
of a string in the class of backgrounds given in (\ref{igeom},
\ref{imetpar}). For a given string frame metric $g_{\mu\nu}(X)$ 
and dilaton $\Phi(X)$ the
covariant Polyakov lagrangian density is given by
\ben
{\mathcal L}_\text{pol} = -\frac{1}{2}{\sqrt{-h}}h^{ab}\partial_aX^\mu 
\partial_b X^\nu~g_{\mu\nu}(X) + {\sqrt{-h}}R~\Phi(X),
\label{lctwo}
\een
where $h_{ab}$ is the worldsheet metric with signature $(-1,1)$.
The string frame
metric which follows from eqs.~\eqref{igeom} and
\eqref{imetpar} may be rewritten as
\ben
ds^2 = e^{\Phi/2} \left[ \frac{e^{f(x^+)}}{Y^2}[2dX^+dX^- + d{\vec{X}}^2]
+ \frac{1}{Y^2}d{\vec{Y}}^2 \right],
\een
where ${\vec{X}}= X^1, X^2$ and ${\vec{Y}}= Y^1 \cdots Y^6$ and
$Y=\abs{\vec{Y}}$ is related to $r$ of (\ref{igeom}) by $Y = R^2/r$.

We will fix a light cone gauge following 
\cite{Metsaev:2000yf,Metsaev:2000yu,Polchinski:2001ju}
\ben
h_{01}=0, \qquad X^+ = \tau.
\label{lctwoa}
\een
Let us first ignore the term containing the dilaton. Then the gauge
fixed action becomes
\ben
{\mathcal L}_\text{pol}
= -\frac{1}{2} \left[ E(\tau,\sigma)~g_{ij}\partial_\tau X^i
\partial_\tau X^j- \frac{1}{E(\tau,\sigma)}g_{ij}\partial_\sigma X^i
\partial_\sigma X^j + 2 E(\tau,\sigma) g_{+-} \partial_\tau X^-  \right],
\label{lcthree}
\een
where we have defined $X^{3\dots 8} = Y^{1\dots 6}$, and
\ben
E(\sigma,\tau) = {\sqrt{-\frac{h_{\sigma\sigma}(\tau,\sigma)}
{h_{\tau\tau}(\tau,\sigma)}}}.
\een
The momentum density conjugate to $X^-$ is
\ben 
{\mathcal P}_- = E (\sigma,\tau) g_{+-}(\tau),
\een
and this is independent of $\tau$ by the equations of motion.
Since the gauge choice (\ref{lctwoa}) still leaves reparametrizations
which are independent of $\tau$, we can choose a $\sigma$ such that
\ben
E (\sigma,\tau) g_{+-}(\tau) = 1.
\een
and use this to write $E$ in terms of $g_{+-}=e^{\Phi/2 + f}/Y^2$. 
The final form of the action is then
\ben
S = \frac{1}{2} \int d\sigma d\tau
 \Bigl[ (\partial_\tau {\vec{X}})^2 +  e^{-f(\tau)}(\partial_\tau
 {\vec{Y}})^2
-\frac{1}{Y^4}e^{2f(\tau)}e^{\Phi(\tau)}(\partial_\sigma {\vec{X}})^2
-\frac{1}{Y^4}e^{f(\tau)}e^{\Phi(\tau)}(\partial_\sigma {\vec{Y}})^2
\Bigr],
\een
where we have to remember that $\tau = X^+$. 
This expression indicates that stringy modes become important when
$e^{f(\tau)}$ and/or $e^{\Phi(\tau)}$ become small. Consider for
example the energy for a piece of a string along the $X^1, X^2$ directions,
at some constant value of ${\vec{Y}} ={\vec{Y}_0} $.
This corresponds to a classical solution, \eg\ $X^1=\lambda \sigma$.
The typical energy, in string units, of this piece at null time
$X^+ = \tau = \tau_0$ is
$E \sim \frac{1}{Y_0^4}e^{2f(\tau_0)}e^{\Phi(\tau_0)} $
which becomes small when the overall factor is small. The factor of
$\frac{1}{Y_0^4}$ is simply the redshift factor of the underlying
geometry.

In the example (\ref{nullsolntanh})  this happens at the 
singularity $X^+ = 0$. Our analysis then suggests that near the
singularity stringy effects become important, in agreement with the
conclusions of the previous section.

In the above analysis we have not considered the effect of the dilaton
coupling to the worldsheet curvature, the second term in
(\ref{lctwo}). However in the gauge we have chosen the intrinsic
worldsheet quantities are functions of $\tau$ and the dilaton is a
function of $\tau$ alone. Therefore this term does not affect the
dynamics of the transverse fields ${\vec{X}},{\vec{Y}}$ and modify the
above conclusion, though the {\em value} of the energy will be
affected.

While it is reasonably clear that stringy effects become important
near the singularity, we have no definitive conclusion about whether
perturbative string theory is well defined  in this background. This would
require a much more detailed study incorporating the RR background in
the fermionic part of the action and a calculation of correlation
functions of vertex operators. We defer this for future work.
However since the string coupling is small and stringy effects are
large near the singularity there is a distinct possibility that 
perturbative string theory is well defined, in which case the
singularity is really resolved by $\alpha^\prime$ corrections.

\section{Penrose Limits and Matrix Theory} \label{sec:matrix}

In this section we perform the Penrose limit of our class of
solutions, eqs.~\eqref{igeom} and~\eqref{imetpar}.
The main motivation behind this is to write
down a Matrix Theory type action which describes the DLCQ 
quantization in the resulting pp-wave background.

\subsection{Penrose Limit} \label{sec:Penrose}

For this purpose it is convenient to rewrite (\ref{igeom}) as
\ben
ds^2 = r^2 [-dt^2+dq^2+e^{F(z^+)}(dx_2^2+dx_3^2)]
+ \frac{dr^2}{r^2} +  d\psi^2 + \sin^2 \psi d\Omega_4^2,
\label{bone}
\een
where we have used the affine parameter $z^+$ along a 
null geodesic instead of the original coordinate 
$X^+ = \frac{1}{\sqrt{2}} (t+q)$
\ben
z^+ = \int^{x^+}dx~e^{f(x)},
\een
and defined a function $F(z^+)$ by
\ben
F(z^+) = f(X^+(z^+))
\een
The coordinates $q,t$ are defined by
\ben
z^+ = \frac{1}{\sqrt{2}} (q+t), \qquad X^- = \frac{1}{\sqrt{2}}(t-q).
\een
In terms of these coordinates the equation (\ref{condnull})
determining
the dilaton becomes
\ben
\frac{1}{2} \left( \frac{d\Phi}{dz^+} \right)^2 + \frac{d^2F}{dz^{+2}} +
\frac{1}{2} \left( \frac{dF}{dz^+} \right)^2 = 0.
\label{bthree}
\een
Now make the following coordinate transformation
\begin{equation}
\begin{split}
r & = \sin~u, \\
t & = -\cot~u -\frac{v}{R^2}+\frac{\phi}{R}, \\
\psi & = \frac{\phi}{R} + u,
\end{split}
\label{btwo}
\end{equation}
as well as rescale
\ben
x^2,x^3  \rightarrow \frac{x^2,x^3}{R}, 
\qquad q  \rightarrow \frac{q}{R}, \qquad
\Omega_4^2  \rightarrow \frac{\Omega_4^2}{R^2}.
\een
Finally we perform the limit
\ben
R \rightarrow \infty, \qquad \text{and} \qquad u,v,X^i,\Omega_4 = \text{fixed}.
\een
In this limit the $S^4$ decompactifies, whose coordinates we
denote by ${\vec{y}}$ and we get the pp-wave metric
\ben
ds^2 = 2 du~dv+ \cos^2u d\phi^2 +
\frac{1}{(G(u))^2} d{\vec{x}}^2 
+\sin^2u(dq^2+d\vec{y}^2),
\label{bsix}
\een
where we have defined 
\ben
G(u) = \frac{e^{-\frac{1}{2} F(-\frac{1}{\sqrt{2}}\cot u)}}{\sin u}
\label{bseven}
\een
This metric may be brought into Brinkmann form by the
coordinate transformations
\begin{equation}
\begin{gathered}
u = U, \qquad
v = V -\frac{1}{2} \xi^2 \tan U +\frac{1}{2} (p^2+\vec{Z}^2) \cot U
    - \frac{1}{2}\frac{\partial_U G}{G} \vec{X}^2, \\
\phi = \frac{1}{\cos U}\xi, \qquad
q = \frac{1}{\sin U}p, \qquad
\vec{y}  =  \frac{1}{\sin U}\vec{Z}, \qquad
\vec{x}  =  \frac{e^{-F/2}}{\sin U}\vec{X},
\end{gathered}
\label{beight}
\end{equation}
[here $G(U) \equiv G(u(U))$] and the pp-wave metric becomes (in Einstein frame)
\ben
ds^2 = 2 dUdV - [H(U)\vec{X}^2 + \vec{Y}^2](dU)^2
+ d\vec{X}^2+d\vec{Y}^2.
\label{bnine}
\een
where we have defined $\vec{Y} = (\vec{Z},p,\xi)$ and the function
$H(U)$ is defined by
\ben
H = \partial_U(\frac{\partial_U G}{G})-(\frac{\partial_U G}{G})^2.
\label{bten}
\een
After some algebra this may be written in terms of the original
function $F(z^+)$ where \hbox{$z^+ = -\frac{1}{\sqrt{2}}\cot U$}
\ben
H(U) = 1 -
\frac{[1+2(z^+)^2]^2}{4} \left[ 
\frac{d^2F}{(dz^+)^2}+\frac{1}{2}\bigl(\frac{dF}{dz^+}\bigr)^2 \right]
= 1+\frac{[1+2(z^+)^2]^2}{8}\left( \frac{d\Phi}{dz^+} \right)^2,
\label{beleven}
\een
where we have expressed this in terms of the dilaton $\Phi (U(z^+))$.
In addition there is a five form field strength which becomes, in
the Penrose limit
\begin{equation} \label{F5}
F_5 = dU \wedge d\xi \wedge dp \wedge dX^2 \wedge dX^3
+ dU \wedge dZ^1 \wedge dZ^2 \wedge dZ^3 \wedge dZ^4.
\end{equation}

It is interesting to examine the nature of the function $H(U)$ for
some specific backgrounds considered in the previous sections.
Consider for example the background given by (\ref{excf}). In this case
\ben
z^+ = X^+ - \tanh X^+.
\een
To obtain the form of the function $H(U)$ we need to invert $X^+$ and
express this as a function of $z^+$ and so obtain $F(z^+)$. Finally we have
to substitute $z^+ = -\frac{1}{\sqrt{2}} \cot U$
and calculate various quantities. Let us 
examine the nature of this function near the singularity at $X^+ = 0$.
This corresponds to $z^+ = 0$ and near the singularity $z^+ \sim 
\frac{(X^+)^3}{3}$. Now, the whole range of values of $U$ cover the
range of $z^+$ from $-\infty$ to $+\infty$ multiple number of times. Consider
one such domain $0 < U < \pi$ which covers $-\infty < z^+ < \infty$. It is then
straightforward to see that near the point $z^+ = 0$ which means $U \rightarrow
\pi/2$ we have
\ben \label{singnearsing}
H(U) \sim \frac{1}{(U-\frac{\pi}{2})^2}, \qquad
e^{\Phi (U)} \sim (U-\frac{\pi}{2})^{\frac{\sqrt{8}}{3}}.
\een
Thus the Penrose limit of our original space-time is singular as well.

In fact, it is easy to see by comparing Eq.~\eqref{beleven} (or,
rather, because of the change of coordinates,~\eqref{condnull}) to
Eq.~\eqref{affcurv} that the pp-wave is singular if and only if
the pre-Penrose limit original spacetime is singular.
That the Penrose limit of a singular
spacetime is also singular was demonstrated in far more generality than
this in~\cite{bbop}.  Interestingly the $\frac{1}{U^2}$-type singular
profile seen in~\eqref{singnearsing}
was shown in~\cite{bbop} to be quite typical of 
the Penrose limit of cosmological singularities, and is amenable to
perturbative string theory analysis~\cite{prt}.

\subsection{Matrix Membrane Action}

In the Penrose limit there are spacelike isometries, which may
be made manifest by choosing a different set of coordinates
\cite{Michelson:2002wa} in which the Einstein frame metric becomes%
\begin{multline}
ds^2 = 2dUdV - 4Y^5dY^6dU - [H(U)\vec{X}^2+(Y^1)^2 + \cdots (Y^4)^2](dU)^2
+ d\vec{X}^2+d\vec{Y}^2.
\label{bninea}
\end{multline}
Consider now the above background with both $V$ and $Y^6$ compact
with radii $R_V$ and $R_B$ respectively. The usual construction of the
Matrix theory dual of the sector of the theory with momentum
$P_V = J/R_V$ along $V$ involves
\begin{enumerate}

\item{} A T duality along $Y^6$ to obtain a IIA theory.
\item{} A lift to M theory by introducing a new direction $Y^7$.
\item{} KK reduction of this M theory along $V$ to yield another IIA
theory
\item{} Performing two T-dualities along $Y^6$ and $Y^7$ on this
IIA theory

\end{enumerate}

Then, following the usual DLCQ logic, the dual theory is a 
2+1 dimensional $SU(J)$ Yang-Mills theory living on a torus.
the action of this theory is obtained by following the same steps
as in \eg\ \cite{Michelson:2002wa,dmichelson2}. Here we quote
the final form of the action
\ben
S = \int d\tau \int_0^{2\pi\frac{l_B^2}{R}} d\sigma
\int_0^{2\pi\frac{g_Bl_B^2}{R}} d\rho~{\mathcal L},
\label{mnsix}
\een
where $l_B, g_B$ are the string length and string coupling
of the original IIB theory. The lagrangian density is
\begin{multline}
{\mathcal L} = \Tr \frac{1}{2} \{ [(D_\tau \chi^\alpha)^2
- e^{\Phi(\tau)} (D_\sigma \chi^\alpha)^2 
- e^{-\Phi(\tau)} (D_\rho \chi^\alpha)^2] \\
+ \frac{1}{G_\text{YM}^2}[e^{\Phi (\tau)} F_{\sigma\tau}^2
+ e^{-\Phi(\tau)}F_{\rho\tau}^2- F_{\rho\sigma}^2] \\
 - H(\tau)[(\chi^1)^2 + (\chi^2)^2] - (\chi^3)^2 
\cdots (\chi^6)^2 - 4 (\chi^7)^2 \\ 
+\frac{G_\text{YM}^2}{2} [ \chi^\alpha , \chi^\beta ]^2 
+ 2 i G_\text{YM} \chi^7 [ \chi^5, \chi^6 ]
    + \frac{4}{G_\text{YM}} \chi^7 F_{\sigma\rho}  \},
\label{mmembraneaction}
\end{multline}
where we have defined a new field \hbox{$\chi^\alpha, \alpha = 1, \cdots
7$} and \hbox{$\chi^i = X^i, i=1,2$} while 
\hbox{$\chi^{i+2} =Y^i, i=1 \cdots 5$}.
Along with the steps outlined above, this lagrangian follows by taking $\tau=U$ gauge, and so $\Phi(\tau)$ here is $\Phi(\tau=U(u(z^+(X^+))))$ of
section~\ref{sec:Penrose}.
The Yang-Mills coupling constant is determined in terms of the
quantities of the IIB theory as 
with
\begin{equation}
G_\text{YM} = \sqrt{\frac{RR_B^2}{g_B l_B^4}}.
\end{equation}
Unlike the matrix membrane actions in \cite{dmichelson2}, the
Yang-Mills coupling defined by the cubic and quartic commutator terms in
the action~(\ref{mmembraneaction}) is time independent. 
Naively it appears that so long
as $g_B \ll 1$ the theory is strongly coupled {\em at all
times} and the fields collapse to diagonal fields in a
suitable gauge. Furthermore the radius of the $\rho$ direction
becomes small and the theory reduces to a 1+1 dimensional theory.
This becomes the light cone worldsheet action after a
dualisation of the gauge field in terms of a new scalar $\chi^8$.
The Matrix membrane lagrangian density becomes in this limit
\begin{multline}
{\mathcal L} = \frac{1}{2} \{ [(\p_\tau \chi^\alpha)^2 + (\p_\tau \chi^8)^2
- e^{\Phi(\tau)} 
(\p_\sigma \chi^\alpha)^2 - e^{\Phi(\tau)} (\p_\sigma \chi^8)^2]\\
 - H(\tau)[(\chi^1)^2 + (\chi^2)^2] - (\chi^3)^2 \cdots (\chi^6)^2
+ 4 \chi^7 \p_\tau \chi^8 \}.
\label{mstringaction}
\end{multline}
The details of the dualisation are provided in appendix~\ref{Mappendix}.
Using the procedure of section~\ref{worldsheet} it is
straightforward to see that this is the light cone gauge worldsheet
lagrangian in the background (Einstein frame) metric (\ref{bninea}),
precisely as it should be.

The lagrangian~\eqref{mmembraneaction}
was written in terms of a flat worldvolume metric.  Let us instead introduce
the worldvolume metric $\gamma$, whose line element is
\begin{equation} \label{wvmm}
ds^2 = - d\tau^2 + e^{-\Phi(\tau)} d\sigma^2 + e^{\Phi(\tau)} d\rho^2.
\end{equation}
Then the bosonic part of the
Matrix membrane action can be written (almost) covariantly as
\begin{multline} \label{curvedMMaction}
S = \int d\tau \int_0^{2\pi \frac{\ell_B^2}{R}} d\sigma
\int_0^{2\pi g_B \frac{\ell_B^2}{R}} d\rho \sqrt{-\gamma}
\left\{
-\frac{1}{2 G^2_{\text{YM}}} \gamma^{ac} \gamma^{bd} F_{ab} F_{cd}
-\frac{1}{2} \gamma^{ab} D_a X^\alpha D_b X^\alpha
\right. \\ \left.
- H(\tau)\left[(\chi^1)^2 + (\chi^2)^2\right]
    - (\chi^3)^2 - \dots - (\chi^6)^2 - 4 (\chi^7)^2
\right. \\ \left.
+ \frac{G^2_\text{YM}}{2} \com{\chi^\alpha}{\chi^\beta}^2
+ 2 i G_\text{YM} \chi^7 \com{\chi^5}{\chi^6}
+ \frac{4}{G_\text{YM}} \chi^7 F_{\sigma \rho}
\right\},
\end{multline}
where $a,b,\dots$ are worldvolume indices. Thus the matrix
membrane theory may be considered to live on a curved space,
albeit with time dependent mass terms.
The curved space,~\eqref{wvmm}, is, however, typically singular at $\tau=0$;
for example, the Ricci scalar is
\begin{equation}
R = \frac{1}{2} \Phi'(\tau)^2,
\end{equation}
which, via~\eqref{beleven}, is singular if the pp-wave is singular.  Since
$\sigma=\rho=\text{const}$ is a geodesic, the $\tau=0$ singularity is at
finite affine parameter.

This story is almost exactly the same as that in~\cite{dmichelson2} in
which the Matrix membrane, for the type IIB maximally supersymmetric
pp-wave deformed to a big bang-type singularity by a null dilaton,
lived on a singular worldvolume.  The difference between that work and
this work is that there the metric was conformally flat, but singular;
here the metric is singular but is not
conformally flat.  This is therefore a further extension
of~\cite{csv0506,dasmichel0508} which discussed Matrix {\em strings\/}
which ended up living on Milne space, and for which, therefore, the
``big bang'' singularity turned into an orbifold singularity. 

Indeed,
like~\cite{dmichelson2} but unlike~\cite{csv0506,dasmichel0508}, we
expect that the Matrix membrane theory exhibits mode production.
Note that this is {\em not} particle production in the target space
theory, which we have shown to be absent. Mode production in the
matrix membrane theory is an extension of a similar phenomenon of
mode production on the light cone worldsheet. Here we consider a {\em fixed}
number of strings with a fixed nonzero value of $k_-$. 
Consider for example a single closed string so that the extent of
the worldsheet $\sigma$ direction is $0 \leq \sigma \leq 2\pi l_s^2 k_-$.
The higher
worldsheet modes are higher oscillator modes of this single
string with the same values of $k_i,k_-$. Worldsheet lagrangians
which are explicitly time dependent would naturally evolve a state 
in the oscillator ground state to higher oscillator states with the
same target space momenta. In our matrix membrane theory the
oscillators of $\chi^i$ are labelled by the quantised momenta $(m,n)$
in the $(\sigma,\rho)$ directions respectively, which correspond 
to oscillator states of single $(p,q)$ strings. The presence of
$\tau$-dependent factors in front of the $D_\sigma \chi^i$ and
$D_\rho \chi^i$ then imply, as in ~\cite{dmichelson2}, that if we start
with the oscillator vacuum in the past, the state near the singularity
would be a squeezed state of higher oscillator modes of a $(p,q)$ string.
In~\cite{dmichelson2} higher modes of a pure F string ($n=0$ modes) 
were not produced.
In the present case, excited states of a F string are produced as well.
This is in accord with our analysis of the worldsheet string theory.

The question whether Matrix Membrane theory ``resolves'' the
singularity of the pp-wave background is related to the question 
whether the worldsheet action in this background leads to
a well defined perturbative string theory. We have not yet been
able to address this question directly. However, 
the connection of the pp-wave with a sector of a $3+1$
dimensional gauge theory suggests that there could be a
nonsingular description.

Recall that the IIB pp-wave discussed above with compact $V,Y^6$ is the 
Penrose limit of $AdS_5 \times \frac{S^5}{Z_{M_1}\times Z_{M_2}}$
\cite{Mukhi:2002ck,Bertolini:2002nr,DeRisi:2004bc}. In this limit,
$R, M_1, M_2 \rightarrow \infty$ and the finite radii $R_V, R_B$ are given by 
\ben
R_V = \frac{R^2}{4M_2}, \qquad R_B = \frac{R}{M_1}.
\een 
States in the pp-wave background with finite $P_V$ and $P_6$ descend
from states in the original background with large angular momenta
along the $S^5$.  The Matrix membrane is supposed to provide a
nonperturbative description of string theory in the original $AdS_5
\times \frac{S^5}{Z_{M_1}\times Z_{M_2}}$ in this large angular
momentum sector, and the momentum modes along $\sigma$ and $\rho$
directions are the F-string and D-string oscillators. 

On the other hand, the AdS/CFT correspondence then implies that there
is a usual dual $3+1$ dimensional gauge theory description along the
lines of \cite{Berenstein:2002jq}, which turns out to be a large
quiver with $M_1M_2$ nodes, each having a $U(N)$ gauge theory
\cite{Bertolini:2002nr,DeRisi:2004bc}. The oscillators of the F-string
now appear in this gauge theory as operators with many scalar
insertions.  

It is natural to expect that this chain of correspondences persist
with our time dependent deformation. Correlation functions of gauge
invariant operators in terms of suitably redefined fields are
therefore expected to be nonsingular.  This might indicate that the
theory could be nonsingular when a correct choice of dynamical
variables is made. However we do not know at this moment how to make
such a choice directly in the worldsheet or Matrix Membrane theory.

In fact, the present paper indicates that different holographic
descriptions are useful to analyse what happens to string theory near
such null singularities. For backgrounds of the type analysed in
\cite{csv0506, li0506, berkooz0507, lisong0507, Hikida:2005ec,
dasmichel0508, Chen:2005mg, she0509, Chen:2005bk, Ishino:2005,
robbinssethi0509, KalyanaRama:2005uw, hikidatai0510, Li:2005ai,
Craps:2006xq, dmichelson2, sethi0603, chen0603, ohta0603, nayak0604,
ohta0605, craps0605, nayak0605}, the Matrix String theory or the
Matrix Membrane theory provided a transparent explanation of how null
singularities may be "resolved". For the kind of backgrounds we have
focussed in this paper, the AdS/CFT type of correspondence is more
suitable.

\acknowledgments
We would like to thank Adel Awad, Atish Dabholkar, Avinash Dhar,
 Rajesh Gopakumar, David Kutasov, 
Gautam Mandal, Tristan McLoughlin,
Shiraz Minwalla, Ashoke Sen,  Alin Tirziu, Spenta Wadia
and Xinkai Wu for discussions. S.R.D. would like to
thank Asia-Pacific Center for Theoretical Physics, Pohang for
hospitality during the Focus Program on Liouville, Integrability and
Branes and Korea Institute for Advanced Studies for hospitality during
Prestring Workshop. The work of S.R.D. and J.M. was supported in part
by National Science Foundation grant Nos PHY-0244811 and PHY-0555444
and by a Department of Energy contract No. DE-FG01-00ER45832.
J.M. was also supported in part by a Department of Energy contract
No. DE-FG02-91ER-40690.
S.P.T's research was supported by  the Swarnajayanti Fellowship, Grant No. ZTHSJ XP-104,
Department of Science and Technology, Govt.\ of India.
K.N. and S.P.T. acknowledge support from the DAE Govt. of India, and 
most of all thank the 
people of India for supporting research in String Theory.   

\appendix
\section{The conformally coupled scalar} \label{sec:confscalar}
We consider the light-cone quantisation 
of the  conformally coupled scalar, whose lagrangian is, 
\be
\label{lag3}
S=-\int d^4x \sqrt{-g}\ [\frac{1}{2} (\partial\varphi)^2 + \frac{1}{6} R \varphi^2 ]
\ee
in a background metric, eq.~(\ref{metc}).

Mode expanding the scalar we get,
\begin{multline}
\label{mecf}
\varphi(x)=e^{-f/2} \int d^2k \int^{\infty}_{0} dk_- \frac{1}{ 
\sqrt{(2\pi)^3 2 k_-}} \left[a(k_i, k_-)
e^{-i(k_i x^i+ k_-X^- + \frac{k_i^2}{2 k_-}X^+)} 
\right. \\ \left.
+
a^\dagger(k_i,k_-) e^{i(k_i x^i+k_-X^- + \frac{k_i^2}{2 k_-}X^+)}\right]
\end{multline}

The momentum conjugate to $\varphi$ is $2 e^f \partial_{X^-}\varphi$. 
One can see that if the creation and annihilation operators satisfy 
the standard commutation relation,
\be
\label{si}
[a(k_i,k_-), a^\dagger(k'_i,k'_-)]=2\pi^3 \delta^2(k_i-k'_i) \delta(k_--k'_-),
\ee
then $\varphi$ satisfies the standard comutation relation with its conjugate 
momentum. 
   
The conformal vacuum is defined as the state which satisfies
the condition~\cite{BD},   
\be
\label{vac}
a(k_i,k_-)|0\rangle=0.
\ee

\section{Calculation of the Bulk Two-Point Function} \label{sec:twopoint}

The equation of motion for a minimally coupled massive scalar in
the metric (\ref{5dimmetric}) is given by
\be
\label{eqma}
\frac{1}{\sqrt{-g}}\del_{\mu}(\sqrt{-g}g^{\mu\nu}\del_{\nu}\varphi) - m^2\varphi = 0.
\ee
Solutions can be found using the method of separation of variables. 
Substituting the ansatz, 
\be
\label{soln}
\varphi(z,x^{\mu})=e^{g(X^+)} e^{ik_-X^-+ik_ix^i} \zeta(z),
\ee
we find that  $g(X^+)$, and $\zeta(z) $ satisfy the equation,
\be
e^{-f}[-ik_-(2g'+f')-k_i^2] - \frac{m^2}{z^2} + \frac{\zeta''}{\zeta} -
\frac{3\zeta'}{z\zeta} = 0.
\ee
Setting the first ($z$-independent) expression equal to a constant 
$-\omega^2$, give,
\be
\label{sepb}
z^2\zeta''-3z\zeta'+(-\omega^2z^2-m^2)\zeta=0, \qquad
-ik_-(2g'+f')-k_i^2 = -\omega^2e^f.
\ee
The first equation for the $z$-part is the same as in $AdS_5$ and its 
solution are Bessel functions, whose solutions are given by (\ref{besselsol}). 
The second equation in eq.~(\ref{sepb}) for $g$ can be solved easily,
yielding the final solution (\ref{finalsol}).

Before proceeding we note that the modes, eq.~(\ref{finalsol}), satisfy the 
completeness relation, 
\begin{multline}
\label{comp}
\int d^4x e^{2f}\ \frac{u_{(k_i,k_-,\omega)}(z=\epsilon, x^\mu)}{ 
\zeta_\omega(\epsilon)}
\frac{ u_{(k_i', k_-',\omega')}(z=\epsilon, x^\mu)}{\zeta_\omega(\epsilon)}
\\= (2\pi)^4 \delta(k_-+k_-')\ \delta^2(k_i+k_i')\ 
\delta(\omega^2-{\omega'}^2)\ |2k_-|.
\end{multline}

\subsection{Calculation from boundary action}

Denote 
$\int d{\vec k}=  \int_{-\infty}^\infty d^2k_i \int_{-\infty}^\infty dk_-
\int_{-\infty}^\infty \frac{d\omega^2}{|2k_-|}$ .
Substituting eq.~(\ref{modeexp}) in the action, eq.~(\ref{scalaraction}), and 
using the equation of motion eq.~(\ref{eqma}), the bulk 
action reduces to a boundary term,
\begin{align}
S &= \left.-\int d^4x \sqrt{-g} g^{zz}\ \varphi({\vec x},z)\
\del_z\varphi({\vec x},z)\right|_{z=\epsilon}
\label{scalaract} \\
&= \int d\vec{k}d\vec{k}' \varphi(k_i, k_-, \omega^2)
\varphi(k_i', k_-',{\omega'}^2)
\delta(k_-+k_-') \delta^{(2)}(k_i+k_i') 2|k_-| \delta(\omega^2-{\omega'}^2)
\omega^{2\nu} C(\nu), \label{valscalaract}
\end{align}
where $C(\nu)$ is a coefficient which depends on $\nu$, and we have dropped
the terms that are singular in $\epsilon$ ({\em c.f.\/}~\cite{vijay9808}).
This can be further simplified to yield (\ref{simpli}).

We would like to convert eq.~(\ref{simpli}) into a position space correlator. 
Using the completeness
relation, eq.~(\ref{comp}), we have,
\be
\label{fme}
\varphi(k_i,k_-,\omega^2)=\frac{\epsilon^{-\Delta_-}}{(2 \pi)^4}\int d^4x 
e^{2f} \varphi(x^\mu, \epsilon)
\frac{u(-k_i,-k_-,\omega^2)(\epsilon,x^\mu)}{\zeta_{\omega}(\epsilon)}
\ee
In the discusion below, we denote,
\be
\label{defp}
\varphi(x)\equiv \epsilon^{-\Delta_-} \varphi(x,\epsilon)
\ee
From, eq.~(\ref{finalsol}), eq.~(\ref{modeexp}), we see that $\varphi(x)$ is 
independent of $\epsilon$.

Then, eq.~(\ref{simpli}) can be written as, 
\be
S = \int d^4x d^4x' e^{3f(X^+)/2} e^{3f({X'}^+)/2}\
\varphi({\vec x}) \varphi({\vec x'})\ G({\vec x},{\vec x'}),
\ee
where
\be
G({\vec x},{\vec x'}) = \frac{1}{(2\pi)^8} \int d{\vec k}\ 
e^{-ik_-\Delta X^--ik_i\Delta x^i}\
e^{-\frac{ik_i^2}{2k_-}\Delta X^+}\ e^{\frac{i\omega^2}{2k_-}\Delta\lambda}
\ \omega^{2\nu},
\label{boundcor}
\ee
with $\Delta x^{\mu} = x^{\mu} - {x'}^{\mu}$.
 
The momentum integrals in $G(\vec{x}, \vec{x'})$ can be carried out and  
gives the final action in position space, (\ref{posnspAction}).

\subsection{Calculation using light front operator quantization}

The {\em normalized} solutions of the Klein-Gordon equation are
\ben
\varphi_{\alpha,\vk,k_-} (\vx,x^\pm,z)
= \frac{1}{\sqrt{2(2\pi)^3}} \sqrt{\frac{\alpha}{2k_-}}
[z^{2} J_\nu (\alpha z)]~[e^{-f(x^+)/2}G_{\vk,k_-,\alpha}(x^+)]~e^{i (\vk
  \cdot \vx + k_-x^-)},
\label{ffone}
\een
where
\ben
G_{\vk,k_-,\alpha}(x^+) = \exp[-i\frac{\vk^2+\alpha^2}{
    2k_-}\lambda(x^+) +i\frac{\vk^2}{2k_-}\kappa (x^+)],
\label{ftwo}
\een 
where
\ben
\lambda(x^+)  =  \int^{x^+} dy~e^{f(y)}, \qquad
\kappa (x^+)  = \lambda (x^+) - x^+.
\label{ffthree}
\een
These modes are normalized according to a Klein-Gordon norm on an
$x^+ = \text{constant}$ surface, which is defined as
\ben
(\varphi_1, \varphi_2) = -i\int dz d\vx dx^-~{\sqrt{-g}}g^{+-}
(\varphi_1 \partial_- \varphi_2^\star - \varphi_2^\star \partial_- \varphi_1).
\een
In this norm we have
\ben
(\varphi_{\alpha,\vk,k_-}, \varphi_{\alpha^\prime,\vk^\prime,k_-^\prime})
= \delta^{(2)}(\vk - \vk^\prime)\delta (\alpha - \alpha^\prime) \delta
(k_- - k_-^\prime).
\een
We have the mode expansion
\begin{multline}
\varphi(x^\pm,z,\vx) = \int_0^\infty d\alpha \int_0^\infty d k_-
\int d^2k~
\left[\varphi_{\alpha,\vk,k_-}(x^\pm,z,\vx)~a(\vk,k_-,\alpha) 
\right. \\ \left. + 
\varphi^\star_{\alpha,\vk, k_-}(x^\pm,z,\vx)~a^\dagger (\vk,k_-, \alpha)\right].
\end{multline}
The commutation relations are
\ben
[ a(\vk,k_-, \alpha) , a^\dagger (\vk^\prime, k_-^\prime, 
\alpha^\prime ] = \delta (\alpha - \alpha^\prime)~
\delta^{2} (\vk - \vk^\prime)~\delta(k_- - k_-^\prime).
\een
The Feynman propagator is
\begin{multline}
 G_F(x^\pm,z,\vx;x^{\pm\prime},z',\vx ') =  
 \int_0^\infty d\alpha \int_0^\infty d k_- 
\int \frac{d^2k}{2(2\pi)^3}~ 
\left(\frac{\alpha}{k_-}\right)(z z^\prime)^{2}~
J_\nu (\alpha z) J_\nu (\alpha z^\prime) \\
\times \left[ e^{i \vk
  \cdot (\vx-\vx^\prime)+i k_-(x^--x^{-\prime})}~
e^{-f(x^+)/2}~G_{\vk,k_-,\alpha}(x^+)
e^{-f(x^{+\prime})/2}~G^\star_{\vk,k_-,\alpha}(x^{+\prime})
\theta (x^+-x^{+\prime}) 
\right. \\ \left.
 + e^{-i \vk
  \cdot (\vx-\vx^\prime) - i k_-(x^--x^{-\prime})}~
e^{-f(x^+)/2}~G^\star_{\vk,k_-,\alpha}(x^+)
e^{-f(x^{+\prime})/2}~G_{\vk,k_-,\alpha}(x^{+\prime})
\theta (x^{+\prime}-x^+)\right].
\label{feynmanfnotzero}
\end{multline}
This may be represented as
\begin{multline}
G_F(x^\pm,z,\vx;x^{\pm\prime},z',\vx ')  =  
e^{-\frac{1}{2}[f(x^+) + f(x^{+\prime})]}~(zz^\prime)^2 
\int_0^\infty d\alpha \int_{-\infty}^\infty dk_- \\
\times \int_{-\infty}^\infty
dk_+ \int d^2 k {\mathcal F}(\vec{k},k_\pm,\alpha).
\label{fffour} 
\end{multline}
where
\begin{multline}
{\mathcal F}(\vec{k},k_\pm,\alpha) =
e^{i\vk \cdot (\vx - \vx^\prime)}e^{ik_-(x^--x^{-\prime})}
e^{-ik_+[\lambda(x^+)-\lambda(x^{+\prime})]}~\\
\times e^{i\frac{\vk^2}{2k_-}[\kappa(x^+)-\kappa(x^{+\prime})]}
\frac{\alpha J_\nu(\alpha z) J_\nu (\alpha z^\prime) 
}{2k_+k_- - (\vk^2 + \alpha^2 -i\epsilon)}.
\label{fnzerorep}
\end{multline}
Since $\lambda (x^+)$ is a monotonic function of
  $x^+$, the integral over $k_+$ in (\ref{fnzerorep}) will 
reproduce (\ref{feynmanfnotzero}).

The integral over $\alpha$ in (\ref{fffour}) may be carried out using the
integrals
\begin{equation}
\begin{aligned}
\int_0^\infty d\alpha~\alpha~
\frac{J_\nu (\alpha z) J_\nu (\alpha z^\prime)}{
2k_+ k_- - (\vk^2 + \alpha^2 - i \epsilon)}
& = I_\nu
(pz) K_\nu (p z^\prime),& \qquad z &< z^\prime, \\
& = K_\nu
(pz) I_\nu (p z^\prime),& \qquad z^\prime &< z,
\end{aligned}
\end{equation}
where
\ben
p = {\sqrt{\vk^2 - 2 k_+k_-}}.
\een
Thus, for $z > z^\prime$ the momentum space propagator may be written as
\begin{multline}
G_{F,\vk,k_-} =
e^{-\frac{1}{2}[f(x^+) + f(x^{+\prime})]}~(zz^\prime)^2 
\int_{-\infty}^\infty dk_+
e^{-ik_+[\lambda(x^+)-\lambda(x^{+\prime})]} \\
\times e^{i\frac{\vk^2}{2k_-}[\kappa(x^+)-\kappa(x^{+\prime})]} 
I_\nu (z^\prime p)K_\nu (z p),
\label{fnzerorep1}
\end{multline}
In the limit $z^\prime \rightarrow 0$ with fixed $z$ the function
$I_\nu (pz^\prime) \sim (pz^\prime)^\nu$. The bulk boundary propagator
is then obtained by dividing by the leading $z^\prime$ dependence of
the Feynman propagator. Finally the boundary correlator is obtained by 
taking the limit $z \rightarrow 0$. The result is easily seen to
agree with (\ref{boundcor}).

\section{From Matrix Membrane action to Worldsheet Action} \label{Mappendix}

Let us start with the abelian version of the ``matrix'' membrane
action~\eqref{mmembraneaction}.
We wish to find the Green-Schwarz string action.  To do this, we follow the
usual procedure of first dualizing the gauge field and then employing
dimensional reduction along $\rho$.
For this, we really only care about the sublagrangian,
\begin{equation}
{\mathcal L}_1 = \frac{1}{G_\text{YM}^2}[e^{\Phi (\tau)} F_{\sigma\tau}^2
+ e^{-\Phi(\tau)}F_{\rho\tau}^2- F_{\rho\sigma}^2]
 + \frac{4}{G_\text{YM}} \chi^7 F_{\sigma\rho} - 4 (\chi^7)^2.
\end{equation}
${\mathcal L}_1$ is that part of the Lagrangian which involves $F$, plus
a term that will disappear when we complete a square.

To dualize the gauge field, we add an auxiliary field $\chi^8$,
and write
\begin{multline}
{\mathcal L}_1 = 
\frac{1}{G_\text{YM}^2}\left[e^{\Phi (\tau)/2} F_{\sigma\tau}
  + G_\text{YM} e^{-\Phi(\tau)/2} \p_\rho \chi^8 \right]^2
\\
+ \frac{1}{G_\text{YM}^2} \left[ e^{-\Phi(\tau)/2}F_{\rho\tau}
  - G_\text{YM} e^{\Phi(\tau)/2} \p_\sigma \chi^8 \right]^2
- \frac{1}{G_\text{YM}^2} \left[ F_{\rho\sigma} - G_\text{YM} \p_\tau \chi^8
    \right]^2
\\
 + \frac{4}{G_\text{YM}} \chi^7 F_{\sigma\rho} - 4 (\chi^7)^2
+ (\p_\tau \chi^8)^2 - e^{\Phi(\tau)} (\p_\sigma \chi^8)^2 
- e^{-\Phi(\tau)}(\p_\rho \chi^8)^2.
\end{multline}
Note that the last three terms ensure that the action is actually
linear in $\chi^8$; the equation of motion for $\chi^8$, in fact, is just
the Bianchi identity $dF=0$.  Therefore, we can treat $F$ as an independent
field---the $\chi^8$ equation of motion is solved by taking $F=dA$---and instead
of integrating out $\chi^8$, we integrate out $F$.
Since $F$ appears quadratically, we can integrate out $F$ by solving
its equations of motion and plugging the solutions back into the Lagrangian.
The equations of motion for $F$ are
\begin{subequations}
\begin{align}
F_{\sigma\tau} &= -G_\text{YM} e^{-\Phi(\tau)} \p_\rho \chi^8, \\
F_{\rho\tau} &= G_\text{YM} e^{\Phi(\tau)} \p_\sigma \chi^8, \\
F_{\sigma\tau} &= G_\text{YM} \p_\tau \chi^8 + 2 G_\text{YM} \chi^7.
\end{align}
\end{subequations}
Therefore, ${\mathcal L}_1$ is equivalent to the action
\begin{equation}
\begin{split}
{\mathcal L}_1 &= 0 + 0 - 4 (\chi^7)^2 + 8 (\chi^7)^2 + 4 \chi^7 \p_\tau \chi^8 - 4 (\chi^7)^2
\\ & \qquad
+ (\p_\tau \chi^8)^2 - e^{\Phi(\tau)} (\p_\sigma \chi^8)^2 
- e^{-\Phi(\tau)}(\p_\rho \chi^8)^2\\
&= 4 \chi^7 \p_\tau \chi^8+ (\p_\tau \chi^8)^2 - e^{\Phi(\tau)} (\p_\sigma \chi^8)^2 
- e^{-\Phi(\tau)}(\p_\rho \chi^8)^2.
\end{split}
\end{equation}

Thus, the Matrix membrane action~\eqref{mmembraneaction} reduces to
the string action (\ref{mstringaction})

\end{document}